\documentclass{elsarticle}
\usepackage{stmaryrd}
\usepackage{amsmath, amssymb, graphicx}
\usepackage[utf8]{inputenc}
\usepackage{lipsum}
\usepackage{overpic}
\usepackage{wasysym}
\usepackage[dvipsnames]{xcolor}
\usepackage{subcaption}
\usepackage{siunitx}
\usepackage{breakurl}
\usepackage[version=4]{mhchem}
\usepackage{listings}
\usepackage{minted}
\usepackage{supertabular,booktabs}
\usepackage[colorlinks = true,linkcolor = blue,urlcolor  = blue, citecolor = blue]{hyperref}
\usepackage[acronym,nonumberlist,nogroupskip]{glossaries}
\usepackage[left=1in,right=1in,top=1in,bottom=1in]{geometry}
\usepackage[export]{adjustbox}
\usepackage{listings}
\usepackage{xcolor}
\usepackage{cleveref}
\usepackage{pythonhighlight}

\crefname{chapter}{Chap.}{Chaps}
\Crefname{chapter}{Chapter}{Chapters}
\crefname{section}{Sect.}{Sects.}
\Crefname{section}{Section}{Sections}
\crefname{figure}{Fig.}{Figs.}
\Crefname{figure}{Figure}{Figures}
\crefname{table}{Table}{Tables}
\Crefname{table}{Table}{Tables}
\crefname{volume}{Vol.}{Vols.}
\Crefname{volume}{Volume}{Volumes}
\crefname{equation}{equation}{equations}
\Crefname{equation}{Equation}{Equations}
\crefname{algorithm}{Alg.}{Algs.}
\Crefname{algorithm}{Algorithm}{Algorithms}
\crefname{definition}{Def.}{Defs.}
\Crefname{definition}{Definition}{Definitions}
\crefname{listing}{code listing}{code listings}
\Crefname{listing}{Code Listing}{Code Listings}
\Crefrangeformat{section}{Sections~#3#1#4--#5#2#6}

\definecolor{codegreen}{rgb}{0,0.6,0}
\definecolor{codegray}{rgb}{0.5,0.5,0.5}
\definecolor{codepurple}{rgb}{0.58,0,0.82}
\definecolor{backcolour}{rgb}{0.95,0.95,0.92}

\lstdefinestyle{mystyle}{
    backgroundcolor=\color{backcolour},   
    commentstyle=\color{codegreen},
    keywordstyle=\color{magenta},
    numberstyle=\tiny\color{codegray},
    stringstyle=\color{codepurple},
    basicstyle=\ttfamily\footnotesize,
    breakatwhitespace=false,         
    breaklines=true,                 
    captionpos=b,                    
    keepspaces=true,                 
    numbers=left,                    
    numbersep=5pt,                  
    showspaces=false,                
    showstringspaces=false,
    showtabs=false,                  
    tabsize=2
}

\def\cm{c_\mathrm{m}}

\def\ct{c_{\mathrm{t}}}

\title{FESTIM v2.0: Upgraded framework for multi-species hydrogen transport and enhanced performance}

\author[MIT]{James Dark\texorpdfstring{\corref{mycorrespondingauthor}}}
\author[MIT]{R\'emi Delaporte-Mathurin}
\author[Simula]{J\o rgen S. Dokken}
\author[MIT]{Huihua Yang}
\author[MIT]{Chirag Khurana}
\author[MIT]{Kaelyn Dunnell}
\author[polito]{Gabriele Ferrero}
\author[mephi]{Vladimir Kulagin}
\author[polito]{Samuele Meschini}

\address[MIT]{Plasma Science and Fusion Center, Massachusetts Institute of Technology, Cambridge, MA 02139, USA}
\address[Simula]{Dept. of Numerical Analysis and Scientific Computing, Simula Research Laboratory, Oslo, Norway}
\address[polito]{Politecnico di Torino, Dipartimento di Energia "Galileo Ferraris", Corso Duca degli Abruzzi 24, Turin, Italy}
\address[mephi]{National Research Nuclear University MEPhI, Moscow, 115409, Russian Federation}

\begin{document}

\begin{abstract}
FESTIM is an open-source finite element framework for modelling the transport of hydrogen isotopes in materials. 
It provides a flexible and extensible tool for simulating diffusion, trapping, surface interactions, and other processes that govern hydrogen behaviour. 
This paper presents FESTIM v2.0, a major release that broadens both the physical scope and the software infrastructure of the framework. 
On the physics side, the formulation adopts a modular structure that supports multi-species transport, advanced trapping and reaction schemes, isotope exchange, decay, and advection. 
Interface and boundary conditions have been generalised, and interoperability with external solvers enables multiphysics workflows, including coupling with fluid dynamics and neutron transport codes. 
On the software side, FESTIM v2.0 has been migrated to DOLFINx, the next-generation FEniCS platform, providing improved performance, interoperability, and long-term sustainability. 
Taken together, these advances position FESTIM v2.0 as a versatile platform for investigating hydrogen transport in materials across scientific and engineering applications. 
\end{abstract}


\maketitle

\section{Introduction}
Hydrogen isotope transport plays a central role in the development of fusion energy systems. 
Accurate modelling of diffusion, trapping, and retention processes is critical for predicting tritium inventories, ensuring regulatory compliance~\cite{gilbert_fusion_2024, dorman_options_2023}, and supporting the design of fuel cycles and plasma-facing components~\cite{abdou_physics_2020}. 
Robust modelling is also essential for guiding near-term experimental campaigns and validation of transport properties in materials. 
Hydrogen transport solvers are routinely applied to laboratory-scale experiments such as gas-driven permeation~\cite{montupet-leblond_permeation_2021, esteban_hydrogen_2001, uehara_hydrogen_2015, calderoni_measurement_2008}, which are used to determine transport properties of candidate materials and coatings, including permeation barriers~\cite{aiello_overview_2004, nemanic_hydrogen_2019, utili_design_2022}. 
They are also used to assess conceptual designs for fuel-cycle components such as breeder blankets~\cite{dark_influence_2021, candido_novel_2021, zhao_3d_2020, ferrero_preliminary_2022}, plasma-facing components~\cite{delaporte-mathurin_fuel_2021, hodille_retention_2017, guterl_modeling_2015}, and tritium extraction systems~\cite{utili_triex-ii_2022, venturini_testing_2024, fuerst_source_2023}, as well as in broader hydrogen-handling technologies including hydrogen storage~\cite{zuttel_hydrogen_2004}, nuclear fission systems~\cite{fiorina_development_2022}, and related materials science studies~\cite{franklin_new_2025, marrani_neural_2025, diaz_comsol_2025, garcia-macias_tds_2024}. 
These diverse applications highlight the broad scientific importance of accurate, extensible, and efficient hydrogen transport solvers. 
FESTIM (Finite Element Simulation of Tritium In Materials) was developed to meet these needs in the context of fusion-relevant materials.

FESTIM was first released as an open-source finite element tool tailored to hydrogen transport modelling~\cite{delaporte-mathurin_festim_2024}, built on the legacy version of the FEniCS Project (2019.1)~\cite{alnaes_fenics_2015}. 
Its Python interface and transparent development model enabled rapid uptake and early community contributions. 
However, FESTIM v1 faced two fundamental limitations. 
First, its core dependency (FEniCS 2019.1) has not been maintained since 2019, which has limited FESTIM's long-term sustainability, scalability, and performance. 
Second, its treatment of material interfaces relied on a change-of-variable method~\cite{delaporte-mathurin_influence_2021} whose implementation limited efficiency and flexibility. 
While adequate for small-scale studies, it became increasingly unsuitable for engineering-focused simulations with fine meshes and multiple materials.

The release of FEniCSx, a modern and actively maintained successor optimised for high-performance computing, provided both the necessity and the opportunity for a complete rewrite of FESTIM. 
FEniCSx is organised as a collection of interoperable modules, including DOLFINx~\cite{baratta_dolfinx_2023} as the problem-solving environment, FFCx for form compilation, Basix~\cite{scroggs_basix_2022} for basis function definitions, and UFL~\cite{alnaes_unified_2014} to describe finite element forms. 
Rather than attempting a direct port, FESTIM v2.0 was re-designed from the ground up with a focus on modularity, extensibility, and computational efficiency. 
This redesign removes the technical debt that had accumulated in v1 and provides a cleaner foundation for extending the physics that FESTIM can describe.

The most significant advance in FESTIM v2.0 is the introduction of a framework that enables fully coupled multi-species transport. 
This new capability allows users to model multiple hydrogen isotopes and their interactions, including multi-level occupancy trapping, isotopic exchange, decay, and advection. 
Such features were not possible in FESTIM v1 due to architectural limitations, which restricted extensibility and maintainability through ad-hoc implementations. 
By contrast, the restructured architecture in v2.0 provides a consistent and modular way to introduce new physics, opening the door to studies of more realistic material behaviour and isotope effects across fusion-relevant conditions.
Alongside this development, FESTIM v2.0 introduces more efficient strategies for handling material interfaces. 
The original change-of-variable approach has been re-implemented to ensure continuity with legacy studies, but FESTIM v2.0 also integrates discontinuous Galerkin formulations, including penalty methods~\cite{babuska_finite_1973, barrett_finite_1986} and Nitsche's method~\cite{hansbo_nitsches_2005, juntunen_nitsches_2009}.
These approaches, well established in the numerical analysis community, substantially reduce the computational cost of multi-material simulations while preserving accuracy. 
In addition, work has been done to facilitate the coupling of external modelling tools (OpenMC~\cite{romano_openmc_2015} and OpenFOAM~\cite{noauthor_openfoam_nodate}) in FESTIM problems.
Together, these advances establish FESTIM v2.0 as both a more capable and more efficient solver for hydrogen transport in materials.

This paper presents FESTIM v2.0 as the new reference for the code and its new capabilities. 
\Cref{sec:festim_overview} provides an overview of the software architecture and development philosophy. 
\Crefrange{sec:geometry}{sec:boundary_conditions} describe the mathematical framework, including geometries, governing equations, interface conditions, and boundary condition treatments. 
\Cref{sec:performance} benchmarks performance improvements relative to FESTIM v1, while \Cref{sec:multiphyics_coupling} demonstrates interoperability with external solvers in multiphysics workflows. 
Finally, \Cref{sec:conclusions} summarises the advances and outlines perspectives for further development. 
All examples and data are openly available at \href{https://github.com/festim-dev/FESTIM-v2-review}{https://github.com/festim-dev/FESTIM-v2-review}.

\section{FESTIM overview}
\label{sec:festim_overview}
\subsection{Philosophy and design principles}
FESTIM has been developed as a domain-specific framework for hydrogen isotope transport, rather than a generic finite element solver. 
Its architecture is guided by three principles: \emph{modularity}, enabling new physics to be incorporated without altering the core; \emph{extensibility}, allowing users to adapt or extend functionality for specialised studies; and \emph{sustainability}, ensuring that FESTIM remains a long-term platform for its users. 
This philosophy reflects the lessons learnt during the development of FESTIM v1, where useful physics extensions were often introduced as ad-hoc workarounds that limited maintainability. 
In v2.0, these have been replaced with a clean and systematic structure, providing a robust foundation for scientific research and engineering-scale applications.

\subsection{Core components}
At its foundation, FESTIM builds on the familiar structure of finite element simulations: mesh, governing equations, boundary conditions, and numerical solvers. 
FESTIM v2.0 extends this foundation with a flexible, object-oriented architecture that assigns physics in a modular manner.

A key innovation is the introduction of \pyth{Subdomain} objects. 
Whereas v1 applied physics globally, v2.0 can now formulate solutions on a per-subdomain basis. 
This allows particular processes or governing equations to be restricted to specific regions of the domain. 
Materials are attached to subdomains rather than the other way around, which improves extensibility and avoids hard-coded links between geometry and physics. 
This makes multi-material problems more natural to construct and provides a consistent way to define material-dependent diffusivities, solubilities, or reaction networks.

Diffusivity is expressed by default through Arrhenius laws, but users may also supply custom \pyth{dolfinx.fem.Function}'s. 
This enables fully spatially or temporally varying fields, supporting advanced cases such as heterogeneous materials or turbulent diffusion without modifying the core.

Thermal coupling has also been restructured. 
In v1, the heat equation was embedded directly within the transport solver, whereas v2.0 separates the two.
Heat transfer problems can be solved directly using the new \pyth{HeatTransferProblem} class.
Users may still solve the heat equation internally with a dedicated solver, or instead provide external temperature fields and couple FESTIM with third-party thermal codes. 
This modularity simplifies the architecture while preserving the ability to run fully coupled thermo-diffusion simulations.

Additional flexibility is provided by classes such as \pyth{ImplicitSpecies}, which represent site balances without introducing extra PDEs. 
This supports efficient modelling of trap occupancies alongside explicit mobile and immobile species. 
Reactions and source terms are defined through extensible interfaces that accommodate trapping, isotope exchange, decay, surface recombination, and other phenomena relevant to hydrogen transport.

Finally, FESTIM includes a configurable export system. 
Users can track both raw fields and derived quantities such as fluxes, inventories, or averages over subdomains, ensuring that simulations can be compared directly with experiments or embedded in broader workflows.

\subsection{Extensibility and integration}
The modular class-based design of FESTIM v2.0 ensures that all major components, species, traps, sources, reactions, boundary conditions, and exports are implemented as extensible classes. 
Users can create custom subclasses to introduce new physics, define non-standard boundary conditions, or extend export routines, without modifying the core code. 
This resolves one of the main limitations of FESTIM v1, where extensions often relied on modifying internal routines, leading to fragile and inconsistent workflows.  
The v2.0 design enables FESTIM to integrate into broader workflows, such as parametric optimisation, multiphysics coupling, or system models.

\subsection{Community and sustainability}
FESTIM is openly developed on GitHub using Git version control~\cite{chacon_pro_2009}. 
The repository supports issue tracking, pull requests, and collaborative review, ensuring contributions are transparent and reproducible. 
This open model has enabled FESTIM to grow into a multi-institutional project, with contributions from over a dozen researchers across academia and national laboratories.  

Modern software practices further reinforce sustainability. 
An extensive test suite is integrated into FESTIM's Continuous Integration (CI) pipeline, automatically verifying new contributions at both unit and system levels. 
Documentation is openly available at \url{https://festim.readthedocs.io}, including theoretical background, a user guide, an API reference, and a contributor's guide. 
Complementary hands-on tutorials are maintained at \url{https://festim-workshop.readthedocs.io}, helping new users explore FESTIM's capabilities. 
Stable releases are distributed via \href{https://pypi.org/project/FESTIM/}{PyPI} under the permissive Apache-2.0 licence, while development versions can be installed directly from source for early access.

\subsection{Verification and validation}
FESTIM v2.0 includes a dedicated and continuously updated verification and validation (V\&V) resource, hosted as a dynamic Jupyter Book at \url{https://festim-vv-report.readthedocs.io}. 
Unlike static publications, this format evolves alongside the code: new benchmark cases can be added as capabilities grow, corrections can be issued immediately, and results remain reproducible and citable. 
This living V\&V record strengthens confidence in FESTIM as a research platform and ensures that its development is directly aligned with experimental validation.

\section{Geometry, meshes and subdomains}
\label{sec:geometry}
\begin{figure}[h!]
    \centering
    \includegraphics[width=\linewidth, trim={2cm 2.5cm 3.5cm 2.5cm}, clip]{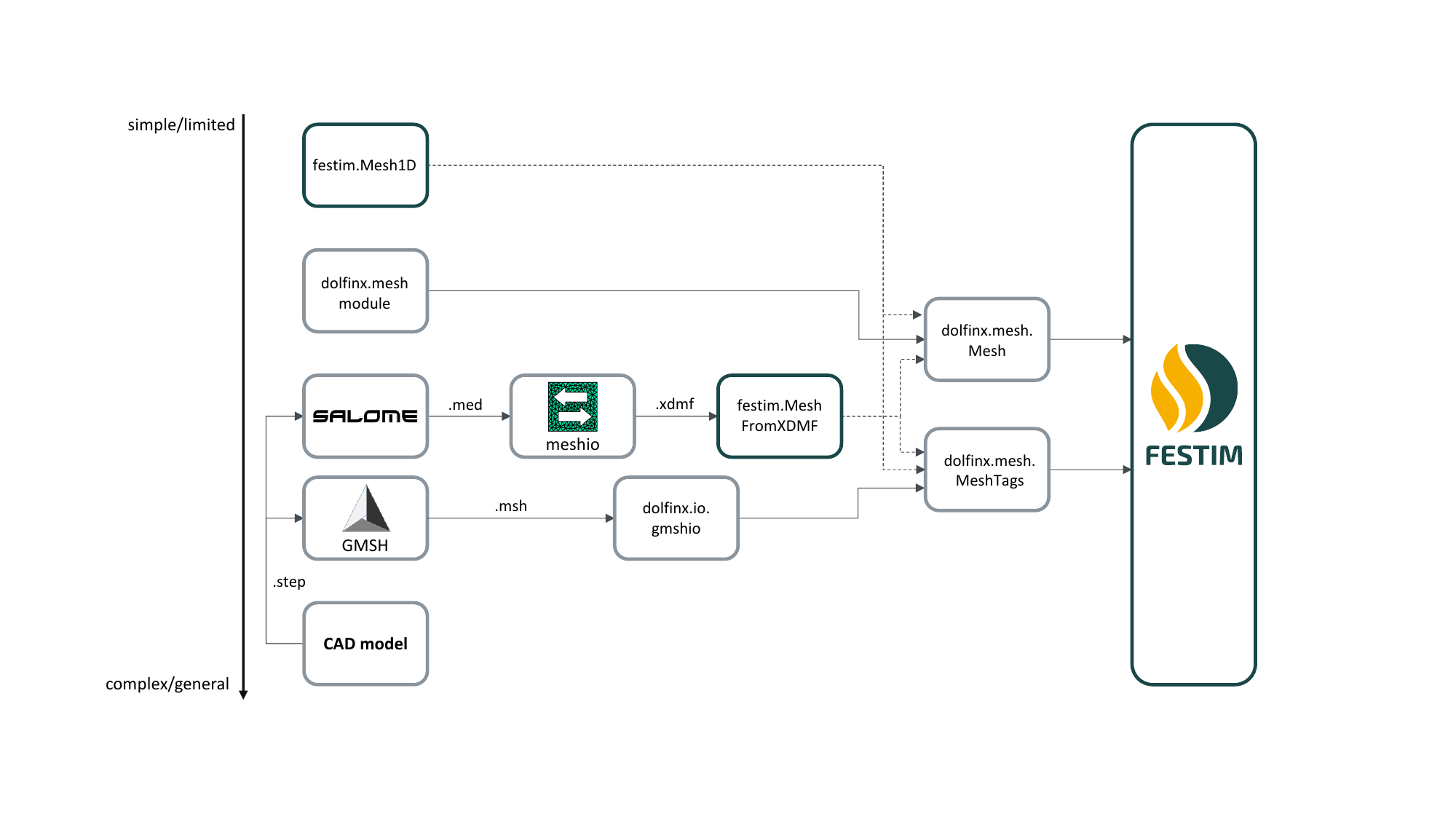}
    \caption{Illustration of how external meshes can be generated and integrated in FESTIM. A user can interact with FESTIM through any of the supported input formats in DOLFINx. Additionally, meshio can be used to convert files from other formats to XDMF, which can then be read into FESTIM.}
    \label{fig:meshing_diag}
\end{figure}

In FESTIM, geometries are represented by a mesh that defines the discretised spatial domain together with mesh tags that identify distinct regions and boundaries. 
These tags enable users to assign materials, sources, or boundary conditions with flexibility. 
All cell types available in DOLFINx are supported, including higher-order triangles, tetrahedra, quadrilaterals, and hexahedra, enabling both simple structured meshes and curved geometries to be modelled with high fidelity (see \cref{fig:meshing_diag}). 
Depending on the problem, users may choose between lightweight built-in options for 1D cases, native DOLFINx constructors for regular 1D-3D domains, or external tools for complex geometries.

For simple 1D studies, FESTIM provides a dedicated \pyth{Mesh1D} class. 
The user specifies vertex coordinates and optional locators for boundaries, and FESTIM constructs the mesh and associated tags automatically:  

\begin{python}
import festim as F
import numpy as np
vertices = np.linspace(0, 1, num=100)

my_mesh = F.Mesh1D(vertices=vertices)
\end{python}

More general meshes can be created directly with DOLFINx constructors such as \pyth{create_interval}, \pyth{create_rectangle}, or \pyth{create_box}. 
Subdomains and boundaries may then be defined using locators or geometric conditions. 
This provides an efficient workflow for simple problems:

\begin{python}
import festim as F
import numpy as np

left_boundary = F.SurfaceSubdomain(id=1, locator=lambda x: np.isclose(x[0], 0.0))
right_boundary = F.SurfaceSubdomain(id=2, locator=lambda x: np.isclose(x[0], 1.0))
bottom_boundary = F.SurfaceSubdomain(id=3, locator=lambda x: np.isclose(x[1], 0.0))
top_boundary = F.SurfaceSubdomain(id=4, locator=lambda x: np.isclose(x[1], 1.0))
\end{python}

For more complex geometries, FESTIM integrates smoothly with meshing tools such as GMSH~\cite{geuzaine_gmsh_2009}, fTetWild~\cite{hu_fast_2020}, or SALOME~\cite{noauthor_salome_2023}, which provide advanced capabilities for generating structured or unstructured meshes and support the import of CAD models. 
Meshes created in SALOME can be exported as \texttt{.med} files and converted to \texttt{.xdmf} with \texttt{meshio}~\cite{schlomer_meshio_2024}, then loaded directly into FESTIM through the \pyth{MeshFromXDMF} class. 
Alternatively, GMSH meshes in \texttt{.msh} format can be converted natively within DOLFINx into \pyth{dolfinx.mesh.Mesh} and \pyth{dolfinx.mesh.MeshTags} objects, which FESTIM can use via the \pyth{Mesh} class. 
Within these frameworks, users can either construct meshes manually or generate them from CAD models, enabling the direct simulation of realistic engineering designs with FESTIM.

\section{Governing equations}
\label{sec:governing_equations}
This section presents the mathematical formulation used by FESTIM to simulate hydrogen transport. 
As in any standard finite element method (FEM) problem, the solution process begins with the definition of the spatial domain, followed by governing equations, boundary and interface conditions, and the specification of desired outputs. 
What distinguishes FESTIM v2.0 is its ability to flexibly combine these elements across multiple materials and species, enabling efficient and accurate treatment of complex, multi-physics scenarios.

In hydrogen transport problems in FESTIM v2.0, one or more \emph{species} can be defined. 
The following generalised partial differential equation governs the concentration of each species: 
\begin{equation} 
    \frac{\partial c_i}{\partial t} = \nabla\cdot (D_i \nabla c_i) + S_i(\mathbf{x}, t, T) + R(c_0, ..., c_n, \mathbf{x}, t, T) + \mathbf{u} \nabla c_i,
    \label{eq:overall_gov} 
\end{equation} 
where $c_i$ is the concentration of species $i$, and $D_i$ is its diffusion coefficient, which may vary in space and time and can depend on the temperature field $T$. 
The volumetric source term, $S_i(\mathbf{x},t,T)$, represents external injection or removal of the species, independent of its interaction with other species. 
The coupling term, $R(c_0, \dots, c_n, \mathbf{x}, t, T)$, accounts for all local interactions between species, such as trapping and detrapping, isotope exchange, and radioactive decay. 
This can act as either a sink or a source, depending on the direction of the interaction, and may exhibit non-linear behaviour in species concentrations. 
The final term, $\mathbf{u} \nabla c_i$, represents advection of the species by a prescribed velocity field $\mathbf{u}$. 
This formulation allows FESTIM to represent independent diffusion processes, externally imposed sources, and complex coupled reaction networks within a unified framework.

It should be noted that this formulation can be further extended to account for additional phenomena, such as thermophoresis (Soret effect)~\cite{dasgupta_impact_2023-1} or stress-assisted diffusion~\cite{reddy_phase-field_2025}.

\subsection{Multi-species}
\label{subsec:multispecies}
One of the new features of FESTIM v2.0 is the support of multiple mobile species.
Species can be classified into two categories: \emph{mobile} and \emph{immobile}.
Mobile species can diffuse through the material, while immobile species cannot, i.e. $D_i$ is zero.

In specific cases, it is possible to define a species as \textit{implicit}, meaning that its concentration can be directly obtained from other species' concentrations.
For instance, for immobile empty trapping sites, for which the concentration $n_\mathrm{empty}$ can be expressed as:

\begin{equation}
    n_\mathrm{empty} = n_\mathrm{total} - c_\mathrm{t},
\end{equation}
where $n_\mathrm{total}$ is the total concentration of trapping sites, and $c_\mathrm{t}$ is the concentration of trapped particles.

\begin{python}
trapped_H = F.Species("ct", mobile=False)
n_total = 1e10

empty_traps = F.ImplicitSpecies(n=n_total, others=[trapped_H])
\end{python}

Here, we assume that a single trapping site can only retain one atom; however, this assumption can be extended to multi-occupancy trapping.
This new paradigm, in combination with the reaction framework, enables users to simulate the transport of multiple hydrogen isotopes and non-hydrogenic species (e.g., vacancies, vacancy clusters, interstitials).

\subsection{Reactions}
\label{subsec:reactions}
One of the key additions to FESTIM v2.0 is the introduction of \textit{reactions}.
Many physical processes in hydrogen transport can be described as reactions. 
A reaction is defined by specifying reactants and products, along with forward and backwards rate constants:

\begin{equation}
    \ce{A + B <-->[k][p] C + D}.
    \label{eq:trapping/reaction_exemple}
\end{equation}

Taking into account only the reaction described in \cref{eq:trapping/reaction_exemple}, the temporal evolution of species concentrations $c_i$ is given by:

\begin{equation}
    \frac{\partial c_\mathrm{A}}{\partial t} \equiv \frac{\partial c_\mathrm{B}}{\partial t} \equiv -\frac{\partial c_\mathrm{C}}{\partial t} \equiv -\frac{\partial c_\mathrm{D}}{\partial t} \equiv -kc_\mathrm{A} c_\mathrm{B} + p c_\mathrm{C} c_\mathrm{D}.
\end{equation}

This framework can be used to model phenomena such as trapping/detrapping, decay, and isotopic exchange.
Reactions can also be combined to create complex reaction schemes.

\subsubsection{Trapping}
The following reversible reaction can represent hydrogen trapping:
\begin{equation}
    \ce{[\quad ] + H <-->[k][p] [H]},
    \label{eq:trapping/detrapping}
\end{equation}
where H is mobile hydrogen, $[\quad ]$ is an empty trap site, and [H] is a trapped hydrogen particle. 

Considering a system with a single mobile species, $\cm$ and a single trapped species, $\ct$, once the user has defined the reactions, the equations are assembled into rate equations.
The temporal evolution of this system is then given by:
\begin{align}
    \frac{\partial \cm}{\partial t} &=  \nabla \cdot (D \nabla \cm)  - \frac{\partial \ct}{\partial t}
    \label{eq:mobile},\\
    \frac{\partial \ct}{\partial t}&=k \ \cm \ (n - \ct)- p \ \ct.
    \label{eq:trapped}
\end{align}
where $n$ is the number of trapping sites.
This is commonly known as the McNabb and Foster model \cite{mcnabb_new_1963}.

The rates of trapping and detrapping, $k$ and $p$ respectively, can be expressed by Arrhenius laws \cite{alefeld_hydrogen_1978}:
\begin{align}
    k &= k_{0} \exp \left( \frac{-E_{k}}{k_\mathrm{B}T} \right),
    \label{eq:trapping_rate}\\
    p &= p_{0} \exp \left( \frac{-E_{p}}{k_\mathrm{B}T} \right),
    \label{eq:detrapping_rate} 
\end{align}
where $k_{0}$ is the trapping pre-exponential factor in \si{m^{3}.s^{-1}}, $E_{k}$ is the trapping energy in \si{eV}. 
$p_0$ is the detrapping pre-exponential factor in \si{s^{-1}}, $E_{\mathrm{p}}$ is the detrapping energy in \si{eV}, \( k_\mathrm{B} \) is the Boltzmann constant in \si{\electronvolt\per\kelvin}.

\begin{python}
mobile_H = F.Species("H")
trapped_H = F.Species("trapped_H", mobile=False)

empty_traps = F.ImplicitSpecies(n=n, others=[trapped_H])

trapping_reac = F.Reaction(
    reactant=[mobile_H, empty_traps],
    product=[trapped_H],
    k_0=k_0, E_k=E_k,
    p_0=p_0, E_p=E_p,
)
\end{python}

\subsubsection{Multilevel trapping}
FESTIM supports multilevel trapping, allowing multiple individuals of the same or different species to be trapped at a single site ~\cite{hodille_study_2016-1}. 
For instance, mobile hydrogen interacting with a trap with three levels can be described by the following set of reactions:
\begin{subequations}
    \begin{align}
    \ce{[\quad ] + H &<-->[k1][p1] [H]}, \\
    \ce{[H] + H &<-->[k2][p2] [2H]} ,\\
    \ce{[2H] + H &<-->[k3][p3] [3H]}.
\end{align}
\end{subequations}

\begin{python}
mobile_H = F.Species("H")
trapped_1H = F.Species("H1", mobile=False)
trapped_2H = F.Species("H2", mobile=False)
trapped_3H = F.Species("H3", mobile=False)

empty_traps = F.ImplicitSpecies(n=10, others=[trapped_1H, trapped_2H, trapped_3H])

reac1 = F.Reaction(
    reactant=[mobile_H, empty_traps],
    product=[trapped_1H],
    k_0=k1,
    p_0=p1,
    ...
)
reac2 = F.Reaction(
    reactant=[mobile_H, trapped_1H],
    product=[trapped_2H],
    k_0=k2,
    p_0=p2,
    ...
)
reac3 = F.Reaction(
    reactant=[mobile_H, trapped_2H],
    product=[trapped_3H],
    k_0=k3,
    p_0=p3,
    ...
)
\end{python}

This can also be extended to multiple species (eg, different hydrogen isotopes): 
\begin{subequations}
\begin{align}
    \ce{[\quad ] + H &<-->[k1][p1] [H]}, \\
    \ce{[\quad ] + D &<-->[k2][p2] [D]}, \\
    \ce{[H] + H &<-->[k3][p3] [2H]}, \\
    \ce{[H] + D &<-->[k4][p4] [H D]}, \\
    \ce{[\mathrm{2H}] + H &<-->[k5][p5] [\mathrm{3H}]}, \\
    \ce{[\mathrm{2H}] + D &<-->[k6][p6] [\mathrm{2H\, D}]}, \\
    \ce{[H D] + D &<-->[k7][p7] [\mathrm{H\, 2D}]}, \\
    \ce{[H D] + H &<-->[k8][p8] [\mathrm{2H\, D}]}.
\end{align}
\end{subequations}

\subsubsection{Isotope swapping}

Isotope swapping refers to the exchange of a mobile hydrogen isotope with another isotope that is already trapped:

\begin{equation}
    \ce{[T] + H <-->[k_\mathrm{swap}][p_\mathrm{swap}] [H] + T}.
    \label{eq:isotope_swap}
\end{equation}

\begin{python}
mobile_H = F.Species("H")
mobile_T = F.Species("T")
trapped_H = F.Species("trapped_H", mobile=False)
trapped_T = F.Species("trapped_T", mobile=False)

swapping_reaction = F.Reaction(
    reactant=[mobile_H, trapped_T],
    product=[mobile_T, trapped_H],
    k_0=k_swap,
    p_0=p_swap,
    ...
)
\end{python}

\subsubsection{Other reactions}
The generalised FESTIM reaction framework also allows the inclusion of additional chemical reactions in the domain. 
This capability is particularly relevant in reactive environments, such as molten salt systems (e.g. FLiBe), where hydrogen may partake in chemical transformations.

Radioactive decay can be formulated as a one-way reaction with no products:
\begin{equation}
    \ce{T ->[\lambda] \emptyset}
\end{equation}
with a decay constant $\lambda$ in \si{s^{-1}}.

\begin{python}
T = F.Species("T")

decay_reaction = F.Reaction(
    reactant=[T],
    product=[],
    k_0=decay_constant,
    ...
)
\end{python}

\subsection{Advection}
To model the transport of hydrogen within a fluid, FESTIM allows the inclusion of an advection term:
\begin{equation}
    \frac{\partial \cm}{\partial t} = \nabla \cdot (D \nabla \cm) + \mathbf{u} \cdot \nabla \cm ,
\end{equation}
where $\mathbf{u}$ is the velocity field in units of \si{m.s^{-1}}. 
This formulation captures the influence of a convecting fluid or moving lattice on the spatial distribution of hydrogen species.

Advection can be included directly using the \pyth{AdvectionTerm} class:
\begin{python}
F.AdvectionTerm(
    velocity=u,
    ...
)
\end{python}
where \pyth{u} is a user-defined \pyth{dolfinx.fem.Function}. 
Further details on coupling velocity fields from CFD solvers are provided in \Cref{subsec:cfd2festim}.

To illustrate the impact of advection relative to diffusion, consider a 2D domain $((0,20) \times (0,10))$ with a constant velocity field $\mathbf{u} = [1, 0]$. 
The initial condition consists of a species concentration $c=1$ within a circular region of radius $0.1$ centred at $(4, 5)$, and $c=0$ elsewhere. 
Two diffusion regimes are examined:
\begin{itemize}
    \item $D = 1.5 \ \si{m^{2}.s^{-1}}$: diffusion-dominated ($\mathrm{Pe}<1$),
    \item $D = 0.01 \ \si{m^{2}.s^{-1}}$: advection-dominated ($\mathrm{Pe}>1$).
\end{itemize}

\begin{figure}[h!]
    \centering
     \begin{subfigure}[b]{0.4\textwidth}
         \centering
         \includegraphics[width=\linewidth]{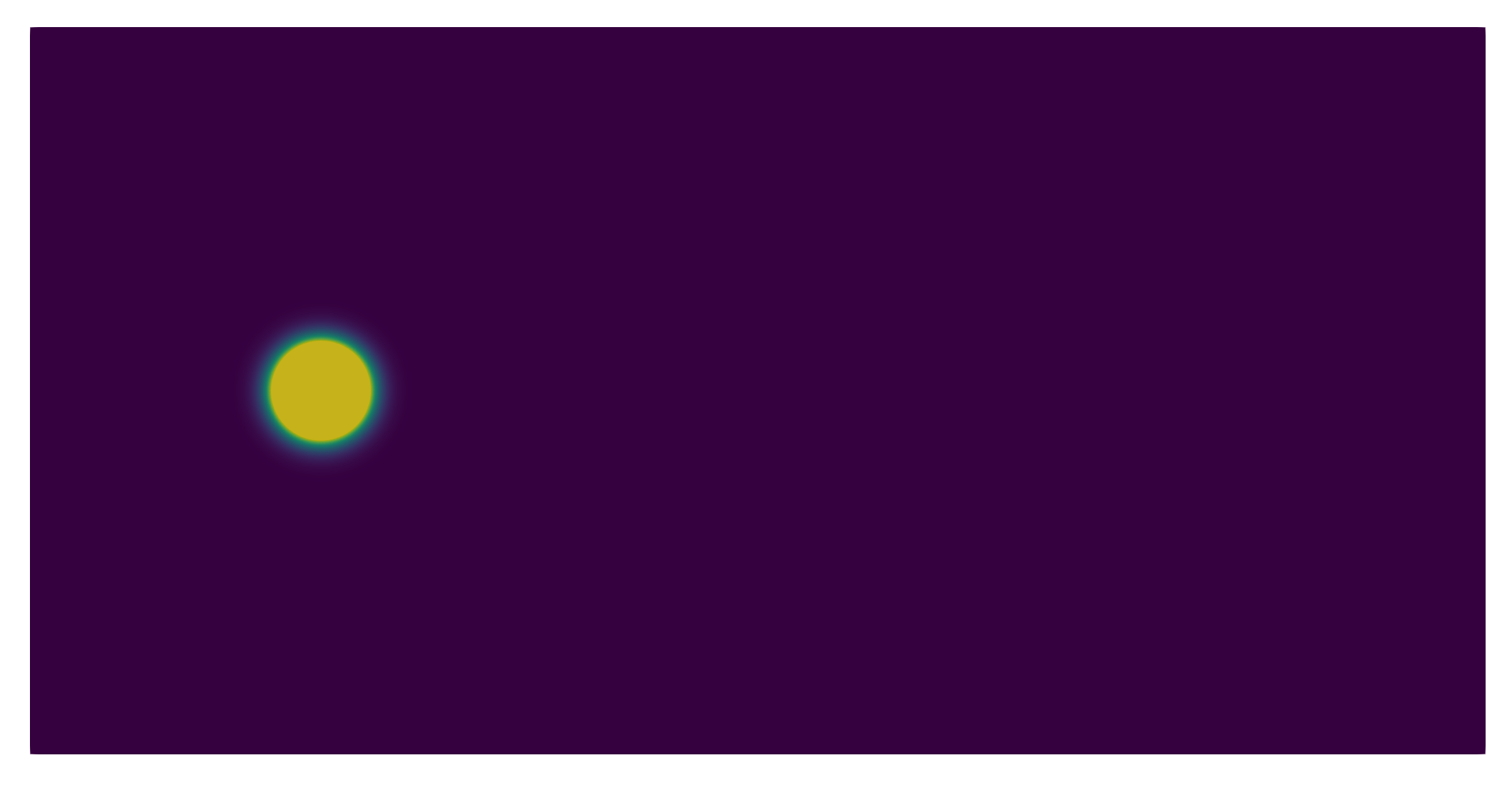}
         \caption{Initial condition, $t=\SI{0}{s}$}
         \label{fig:init_cond}
     \end{subfigure} \\
     \vspace{1em}
     \begin{subfigure}[b]{0.4\textwidth}
         \centering
         \includegraphics[width=\linewidth]{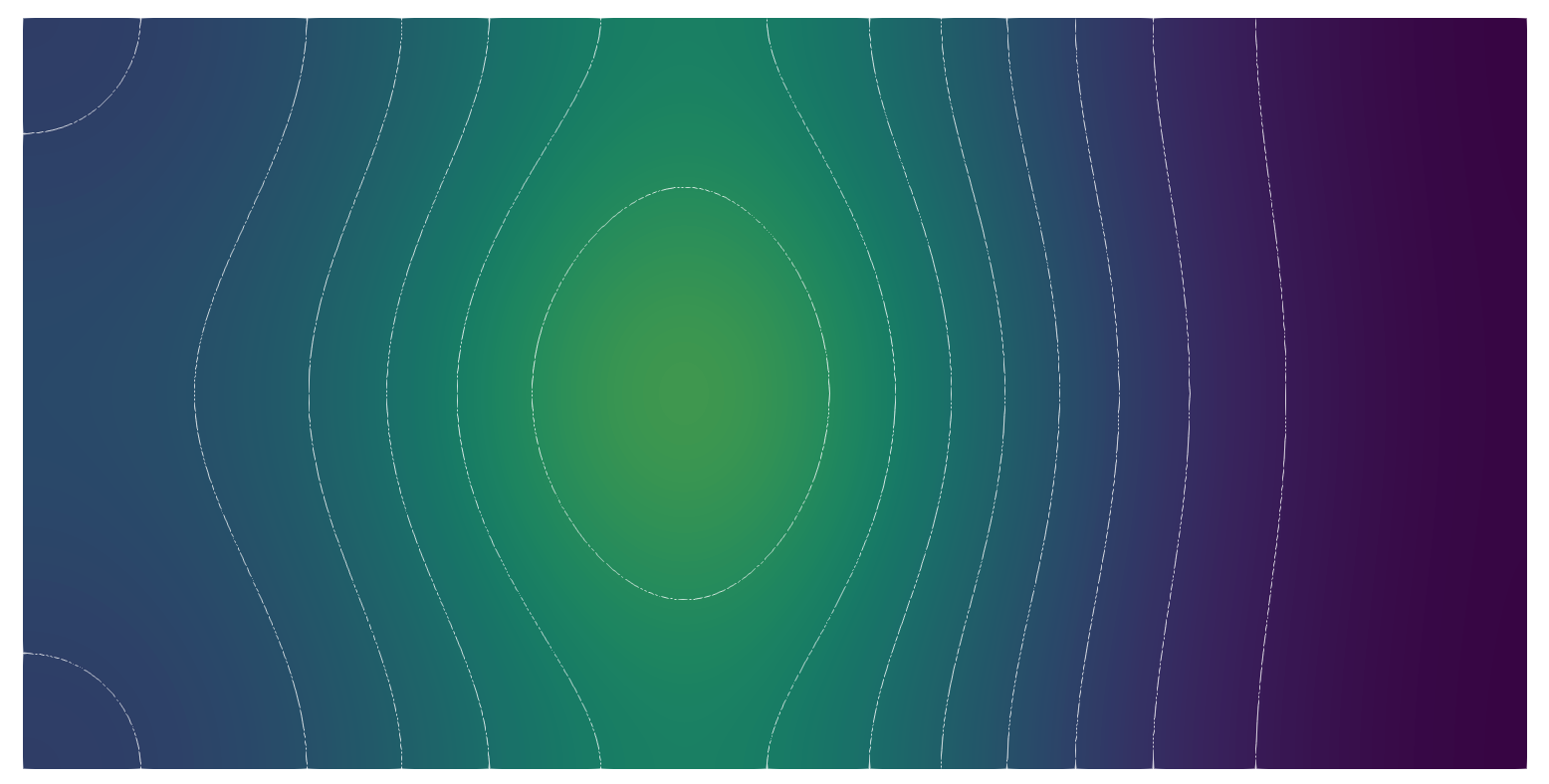}
         \caption{Diffusion dominated, $t=\SI{5}{s}$}
     \end{subfigure}
     \quad \quad
     \begin{subfigure}[b]{0.4\textwidth}
         \centering
         \includegraphics[width=\linewidth]{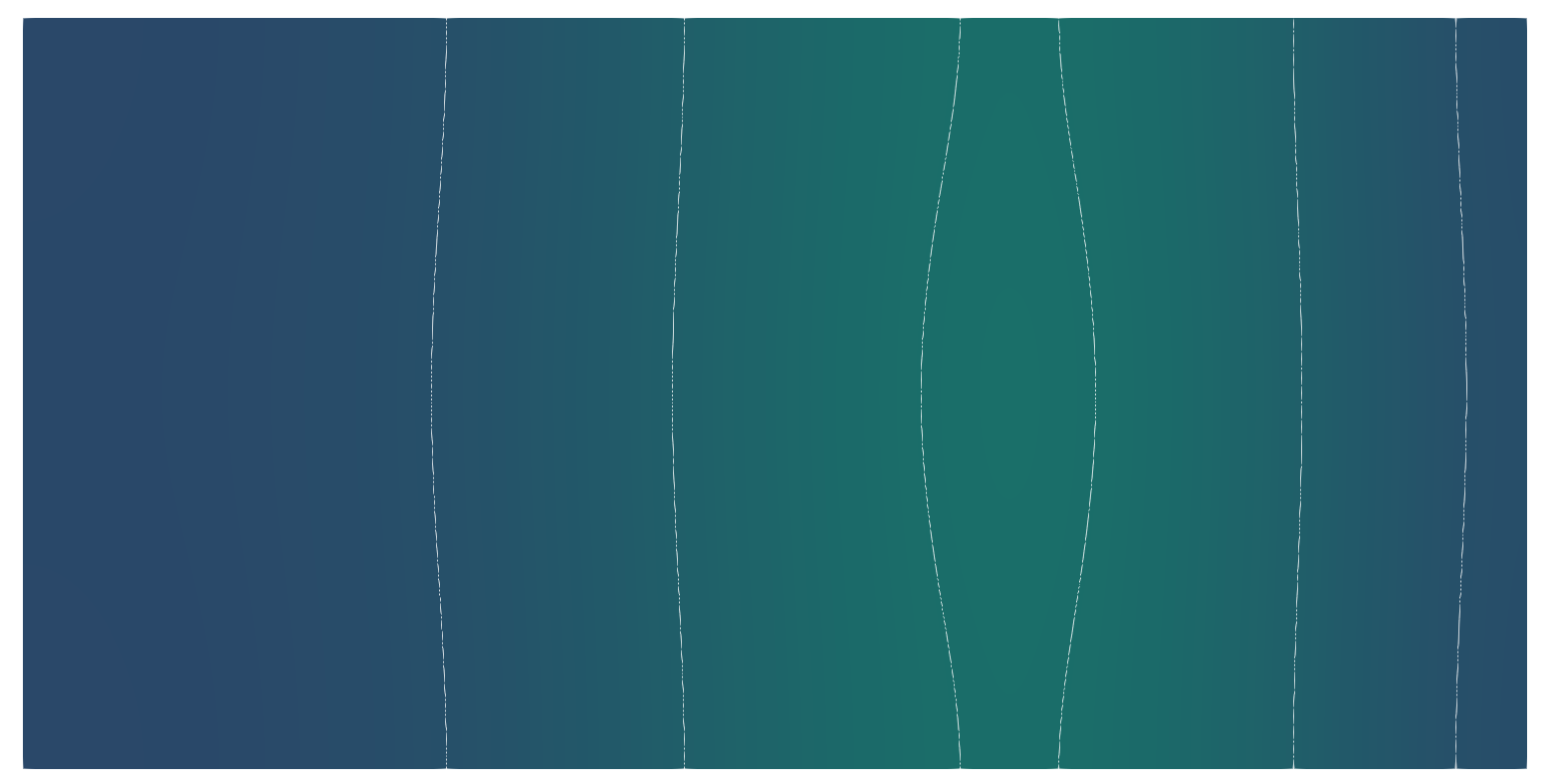}
         \caption{Diffusion dominated, $t=\SI{10}{s}$}
    \end{subfigure} \\
    \vspace{1em}
    \begin{subfigure}[b]{0.4\textwidth}
         \centering
         \includegraphics[width=\linewidth]{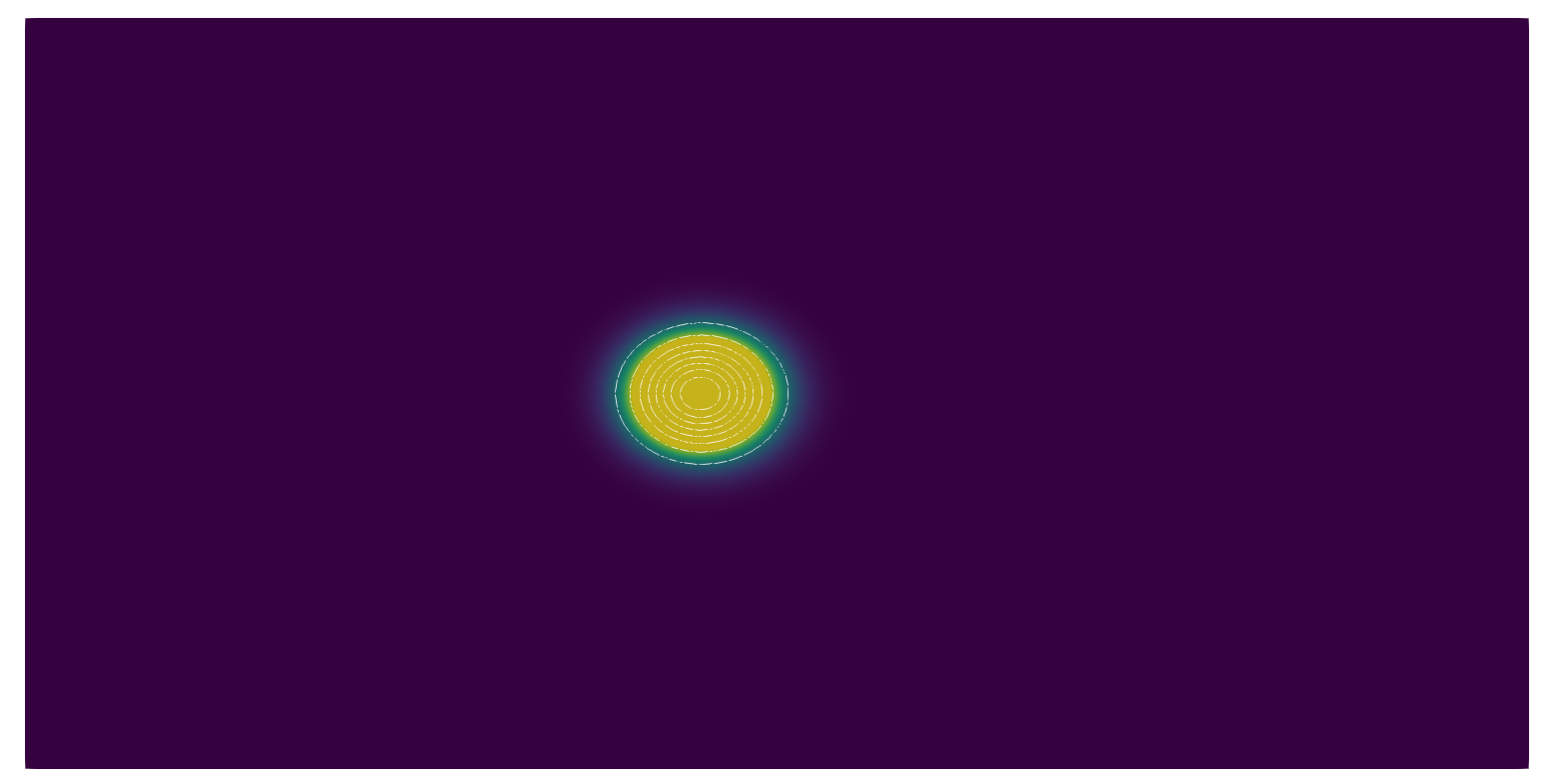}
         \caption{Advection dominated, $t=\SI{5}{s}$}
     \end{subfigure}
     \quad \quad
     \begin{subfigure}[b]{0.4\textwidth}
         \centering
         \includegraphics[width=\linewidth]{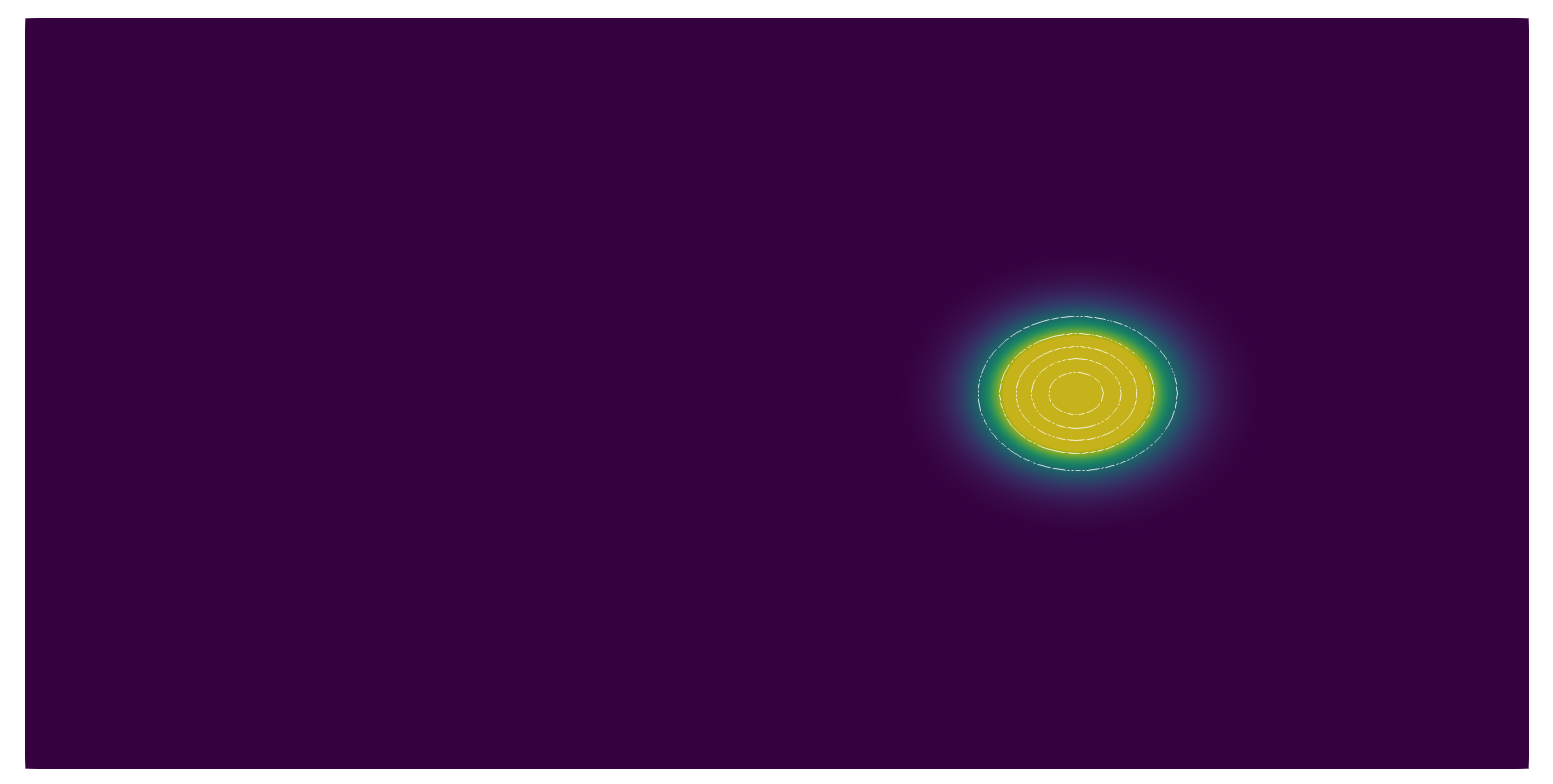}
         \caption{Advection dominated, $t=\SI{10}{s}$}
     \end{subfigure}
        \caption{Influence of advection on species transport in diffusion- and advection-dominated regimes.}
        \label{fig:advection_vis}
\end{figure}

\Cref{fig:advection_vis} shows the concentration fields at $t=\SI{5}{s}$ and $t=\SI{10}{s}$. 
In the diffusion-dominated regime, the species spreads isotropically. 
In contrast, in the advection-dominated regime, the concentration distribution is transported downstream by the flow, retaining its shape. 
Some artificial dissipation of the solution is visible in this case, which is expected when solving the advection-diffusion equation with continuous Galerkin elements in a finite element framework~\cite{siegel_solution_1997}.

\section{Interface conditions}
\label{sec:interface_conditions}
When hydrogen moves through a component composed of multiple materials, it encounters interfaces where transport behaviour can change abruptly. 
The appropriate interface condition depends on the materials involved, reflecting differences in their solubility laws. 
For example, at an interface between two Sieverts materials, continuity of hydrogen partial pressure gives:
\begin{align}
   \frac{c_{\mathrm{m},0}}{K_{\mathrm{S},0}} &= \frac{c_{\mathrm{m},1}}{K_{\mathrm{S},1}},
\end{align}
similarly for two Henry materials:
\begin{align}
   \frac{c_{\mathrm{m},0}}{K_{\mathrm{H},0}} &= \frac{c_{\mathrm{m},1}}{K_{\mathrm{H},1}},
\end{align}
where $K$ is the solubility of hydrogen in a material. 
If the solubility differs across the interface ($K_{\cdot,0}\neq K_{\cdot,1}$), the hydrogen concentration $c_m$ may exhibit a discontinuity.

In FESTIM v1, such conditions were handled by a change-of-variable method, solving for $\theta = \cm/K$ rather than $\cm$. 
Although straightforward, this approach had limitations: it required costly projections during post-processing and prevented species from being restricted to specific subsets of the computational domain.

FESTIM v2.0 introduces several alternative strategies for handling interface conditions, providing users with the flexibility to select the method most suitable for their application. 
Options include discontinuous Galerkin/Nitsche formulations and penalty-based approaches. 
The following subsections outline these approaches.

\subsection{Discontinuous Galerkin/Nitsche method}
FESTIM v2.0 leverages the submesh functionality provided by DOLFINx~\cite{dean_mathematical_2023, richardson_efficient_2025}. 
For interfaces between materials of the same type (either Sievert or Henry), a discontinuous Galerkin/Nitsche method is available \cite{arnold_unified_2002}.
To illustrate, consider two species, $c_{\mathrm{m},i}\in \Omega_i$, $i=0,1$, where $\Omega_0\cap\Omega_1=\Gamma$ denotes their interface. 
The variational formulation of Section \ref{sec:governing_equations} can then be written as

Find $c^n_{\mathrm{m}, i} \in V(\Omega_i)$ such that:
\begin{equation}
    \begin{split}
        &\sum_{i=0}^1\left(\int_{\Omega_i} \frac{c_{\mathrm{m},i}^n-c_{\mathrm{m}, i}^{n-1}}{\delta t} v_i 
        + D \nabla c^n_{\mathrm{m}, i}\cdot \nabla v_i 
        - (\mathbf{u} \cdot \nabla c^n_{\mathrm{m}, i}) v_i~\mathrm{d}x\right) \\
        &- \int_{\Gamma} \langle D \nabla \cm^n\rangle \cdot \mathbf{n} \llbracket v \rrbracket
        - \langle D \nabla v \rangle\cdot\mathbf{n}\left\llbracket \frac{\cm^n}{K}\right\rrbracket 
        + \frac{\gamma}{\langle h\rangle}\left\llbracket \frac{\cm^n}{K} \right\rrbracket\llbracket v \rrbracket ~\mathrm{d}s \\
        &= \sum_{i=0}^1 \int_{\Omega_i}S_i \cdot v_i ~\mathrm{d}x
        \qquad \forall v_i\in V(\Omega_i).
    \end{split}
\end{equation}
Here, $\langle \cdot \rangle:=\frac{\cdot_{0}+\cdot_{1}}{2}$ represents the average operator at the interface, $\llbracket \cdot \rrbracket:=\cdot_{0}-\cdot_{1}$ is the jump operator, and $\mathbf{n} = \mathbf{n}_0$ is the normal vector on $\Gamma$ pointing out of $\Omega_0$.
$\gamma$ is the stabilisation of penalty parameter and $h$ is the local mesh size.
Note that on each subdomain $\Omega_i$, $V(\Omega_i)$ is discretised with a continuous Lagrange space, minimising the number of degrees of freedom needed to represent the discontinuity at $\Gamma$.
This formulation allows the weak enforcement of continuity conditions across the interface while still accommodating possible discontinuities in $\cm$ when material properties differ.

\subsection{Penalty methods}
For more complex cases, such as Sievert-Henry interfaces, the relation between concentrations becomes nonlinear:
\begin{align}
    \left(\frac{c_{\mathrm{m}, 0}}{K_\mathrm{S}}\right)^2 &= \frac{c_{\mathrm{m}, 1}}{K_\mathrm{H}}.
\end{align}
Handling such nonlinear interface conditions requires alternative strategies. 
FESTIM v2.0 offers penalty-based methods that enhance the weak formulation to enforce the interface constraint efficiently. 
Two variants are available: an energy-based formulation and a Robin-type boundary condition.

\subsubsection{Energy perspective}
We consider the same domains and interfaces as above.
The strong formulation of the problem is
\begin{align}
    -\nabla \cdot (D_i\nabla c_\mathrm{m,i})&=S_i&&\text{in } \Omega_i, \quad i=0,1\\
    c_\mathrm{m,0} &= K_0\left(\frac{c_\mathrm{m,1}}{K_1}\right)^n&&\text{on } \Gamma,\\
    D_0\nabla c_\mathrm{m,0} \cdot \mathbf{n}_0 &=  D_1\nabla c_\mathrm{m,1}\cdot \mathbf{n}_0 && \text{on }\Gamma
\end{align}
where $n=\left\{\frac{1}{2}, 1 \right\}$.

The corresponding energy minimisation problem can be expressed as:
\begin{align}
    \min_{c_\mathrm{m,0}, c_\mathrm{m,1}} J(c_\mathrm{m,0}, c_\mathrm{m,1}) \notag \\
    \text{where} \notag \\
    J(c_\mathrm{m,0}, c_\mathrm{m,1}) &=\sum_{i=0}^1
    \int_{\Omega_i}\frac{1}{2}D_i\nabla c_\mathrm{m,i}\cdot \nabla c_\mathrm{m,i} - S_i\cdot c_\mathrm{m,i}~\mathrm{d}x
\end{align}
To enforce the non-linear interface condition, the functional is augmented with a quadratic penalty term:
\begin{align}
    \frac{1}{2}\alpha\int_\Gamma \left(c_\mathrm{m, 0} - K_0\left(\frac{c_\mathrm{m,1}}{K_1}\right)^n\right)^2~\mathrm{d}s
\end{align}
where $\alpha$ is a penalty parameter controlling the strength of the constraint. 
This leads to the following weak formulation by finding the optimality conditions:
Find $c_\mathrm{m,0}\in V(\Omega_0), c_\mathrm{m,1}\in V(\Omega_1)$ such that:
\begin{equation}
    \begin{split}
         &\sum_{i=0}^1\left(\int_{\Omega_i} D_i\nabla c_\mathrm{m,i} \cdot \nabla v_i - S_i v_i~\mathrm{d}x\right) \\
         &\quad + \alpha\int_{\Gamma}
         \left(c_\mathrm{m,0} - K_0\left(\frac{c_\mathrm{m,1}}{K_1}\right)^n\right)
         \left(v_0 - n\frac{K_0}{K_1} \left(\frac{c_\mathrm{m,1}}{K_1}\right)^{n-1}v_1\right)~\mathrm{d}s = 0.
    \end{split}
\end{equation}
This energy-based approach incorporates nonlinear interface conditions into a minimisation problem, which can be solved efficiently in the finite element framework. 
The penalty term enforces the interface constraint weakly, allowing a controlled approximation while maintaining computational stability.

\subsubsection{Change of boundary condition}
An alternative penalty formulation is to impose the condition directly as a Robin-type boundary condition on both sides of the interface:
\begin{align}
    -D_0\nabla c_\mathrm{m,0} \cdot \mathbf{n_0} &= \alpha \left( c_\mathrm{m,0} -  K_0\left(\frac{c_\mathrm{m,1}}{K_1}\right)^n\right)&&\text{on } \Gamma,\\
    -D_1\nabla c_\mathrm{m,1} \cdot \mathbf{n_0} &= \alpha \left( c_\mathrm{m,0} -  K_0\left(\frac{c_\mathrm{m,1}}{K_1}\right)^n\right)&&\text{on } \Gamma
\end{align}

This formulation enforces the interface condition in a weak sense.

Starting from the strong form, multiplication by the corresponding test functions and integration by parts yields the weak formulation
\begin{align}
   \sum_{i=0}^1\left(\int_{\Omega_i} D_i\nabla c_\mathrm{m,i} \cdot \nabla v_i - S_i v_i~\mathrm{d}x\right)
   +\alpha \int_\Gamma \left(c_\mathrm{m,0} - K_0\left(\frac{c_\mathrm{m,1}}{K_1}\right)^n\right)(v_0-v_1)~\mathrm{ds}=0.
\end{align}

\section{Boundary conditions}
\label{sec:boundary_conditions}
FESTIM v2.0 provides a comprehensive set of boundary condition treatments. 
These range from simple fixed concentrations and prescribed fluxes to a general framework for surface reactions, which derives boundary fluxes directly from chemical processes at material interfaces. 
This breadth of functionality allows users to model both imposed concentrations and surface kinetics within the same simulation environment.

\subsection{Fixed concentrations and surface fluxes}
Dirichlet boundary conditions specify the value of a species concentration at a boundary. 
This imposes a constant concentration throughout the simulation at the selected location, which may be used to model perfect sinks, imposed gas-side conditions, or fixed inventory states. 
FESTIM offers several convenience classes to help users define common fixed-value conditions. 
These include options for deriving boundary concentrations from pressure-based relationships such as Sieverts' or Henry's law, where the surface concentration depends on a given gas pressure and material solubility.

More generally, users can prescribe any arbitrary function of space, time, and temperature to define a fixed boundary value. 
This is particularly useful in simulations with time-dependent values or temperature-evolving conditions:
\begin{python}
my_custom_value = lambda x, t, T: 2 * x[0] + 3 * t**2 + 4 * T

my_fixed_bc = F.FixedConcentrationBC(
    value=my_custom_value,
    ...
)
\end{python}

Similarly, flux boundary conditions can be defined using the \texttt{ParticleFluxBC} class. 
Here, the boundary condition enforces a prescribed particle flux across the surface. 
Flux values may depend not only on position, time, and temperature, but also on the concentrations of one or more species:
\begin{python}
H = F.Species("H")
my_custom_value = lambda x, t, T, c_H: x[0] / 2 + t**2 + 2 * T + 10 * c_H

my_flux_bc = F.ParticleFluxBC(
    value=my_custom_value,
    species=H,
    species_dependent_value={"c_H": H}
    ...
)
\end{python}
This flexibility enables a wide range of scenarios to be represented, including time-varying injections, decaying sources, and feedback-driven fluxes that dynamically depend on local concentrations.

\subsection{Surface reactions}
One of the key additions in FESTIM v2.0 is the introduction of surface reactions, a generalised framework for defining species fluxes at boundaries. 
This extends the reaction-based approach described in Section~\ref{subsec:reactions} to surfaces, allowing users to describe a wide range of physical processes through chemical kinetics. 
A surface reaction is defined by specifying the reactants and products, along with forward and backwards rate constants. 
For example, a general surface reaction may be written as:
\begin{equation}
\ce{A + B <-->[k][p] C}.
\end{equation}
Users define such reactions using the \texttt{SurfaceReaction} class:
\begin{python}
A = F.Species("A")
B = F.Species("B")
C = F.Species("C")

react1 = F.SurfaceReaction(
    surface=1,
    reactant=[A, B],
    product=[C],
    k_r0=k,
    k_d0=p,
    ...
)
\end{python}
The resulting flux is evaluated automatically and applied as a Neumann-type boundary condition:
\begin{equation}
    R = k c_\mathrm{A} c_\mathrm{B} - p c_\mathrm{C}.
\end{equation}

This framework can be used to describe recombination and dissociation processes, such as:
\begin{equation}
    \ce{H + H <-->[K_r][K_d] H_2}
    \label{eq:recomb/diss}
\end{equation}
Forward and backwards rates can be expressed with Arrhenius-type laws.
The flux is therefore expressed as:
\begin{equation}
    \mathbf{J}_{\mathrm{H}} \cdot \mathbf{n} = 2(K_r c_{\mathrm{H}}^2 - K_d P_{\mathrm{H_2}})
\end{equation}
Where $P_{\mathrm{H_2}}$ is the partial pressure of $\mathrm{H_2}$ in units \si{Pa}.

Surface reactions can involve multiple isotopes, enabling more complex pathways such as:
\begin{subequations}
    \begin{align}
        \ce{H + H &<-->[K_r1][K_d1] H_2}, \\
        \ce{T + T &<-->[K_r2][K_d2] T_2}, \\
        \ce{H + T &<-->[K_r3][K_d3] HT}.
    \end{align}
\end{subequations}
The corresponding fluxes for H and T can then be expressed as:
\begin{align}
    \mathbf{J}_{\mathrm{H}} \cdot \mathbf{n} &= 2(K_{r1} c_{\mathrm{H}}^2 - K_{d1} P_{\mathrm{H_2}}) + K_{r3} c_{\mathrm{H}}c_{\mathrm{T}} - K_{d3} P_{\mathrm{HT}} \\
    \mathbf{J}_{\mathrm{T}} \cdot \mathbf{n} &= 2(K_{r2} c_{\mathrm{T}}^2 - K_{d2} P_{\mathrm{T_2}}) + K_{r3} c_{\mathrm{H}}c_{\mathrm{T}} - K_{d3} P_{\mathrm{HT}}
\end{align}

Each reaction is declared explicitly by the user, and FESTIM automatically applies the corresponding flux conditions:
\begin{python}
H = F.Species("H")
T = F.Species("T")

reac1 = F.SurfaceReaction(
    surface=1,
    reactant=[H, H],
    k_r0=K_r1,
    k_d0=K_d1,
    ...
)

reac2 = F.SurfaceReaction(
    surface=1,
    reactant=[T, T],
    k_r0=K_r2,
    k_d0=K_d2,
    ...
)

reac3 = F.SurfaceReaction(
    surface=1,
    reactant=[H, T],
    k_r0=K_r3,
    k_d0=K_d3,
    ...
)
\end{python}

Finally, isotopic exchange can also be modelled. 
For example:
\begin{equation}
    \ce{T + H_2 <--> HT + H}.
\end{equation}
In cases where one species dominates (e.g. $P_{\ce{H_2}} \gg P_{\ce{HT}}$), the reverse reaction may be neglected, and the process reduces to a first-order flux in \ce{T}. 
This can be implemented using the \texttt{ParticleFluxBC} class, with a user-defined expression based on local tritium concentration and a fixed \ce{H_2} background.

\section{Performance benchmark}
\label{sec:performance}
To illustrate the benefits of the new implementation in FESTIM v2.0, we consider the 
\textit{Diffusion: multi-material} benchmark case from the FESTIM verification and validation (V\&V) book~\cite{delaporte-mathurin_festim_2024-1}, available at \url{festim-vv-report.readthedocs.io/en/latest/verification/mms/discontinuity.html}. 
This case was selected because it highlights two essential aspects: (i) it directly probes the treatment of material interfaces, which represented a bottleneck in FESTIM v1, and (ii) it has been deliberately configured to run for long physical times on a sufficiently large mesh, thereby exposing performance differences between the available interface treatments. 
The same example has also been shown to exhibit good numerical scaling, as documented on the example's page. 
All input files are openly available.  

\begin{figure}[h]
    \centering
    \includegraphics[width=0.4\linewidth]{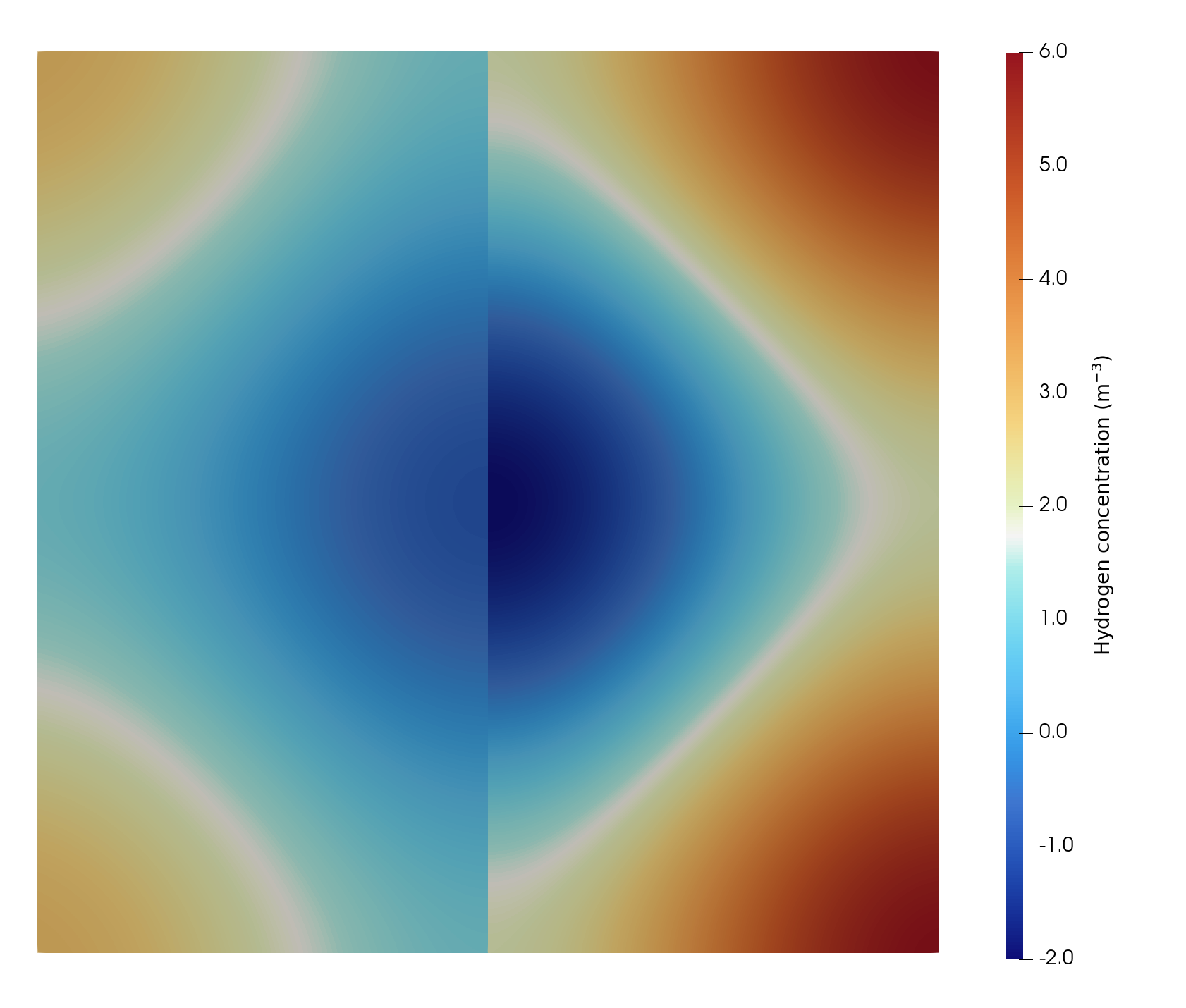}
    \caption{Multi-material test case hydrogen field solution}
    \label{fig:placeholder}
\end{figure}

The benchmark simulates 2D hydrogen diffusion across a multi-material system in the transient mode. 
The simulation was run with a uniform $300 \times 300$ mesh of a unit square, a time step of \SI{0.1}{s}, and a total simulation time of \SI{200}{s}. 
All simulations were executed on a single core of a workstation equipped with an Intel(R) Core(TM) i9-285 K CPU (\SI{3.70}{GHz}) and \SI{191}{GB} of RAM. 
This choice of hardware (a Windows workstation rather than a high-performance computing cluster) was deliberate, as it is representative of the environment in which most FESTIM users currently operate. 
Each configuration was repeated ten times, and the mean runtime was recorded. 
In FESTIM v1, the only available interface treatment was the change-of-variable method. 
FESTIM v2.0 retains this method for continuity with legacy studies and introduces alternative methods (see \Cref{sec:interface_conditions}).

\begin{figure}[h]
    \centering
    \includegraphics[width=0.75\linewidth]{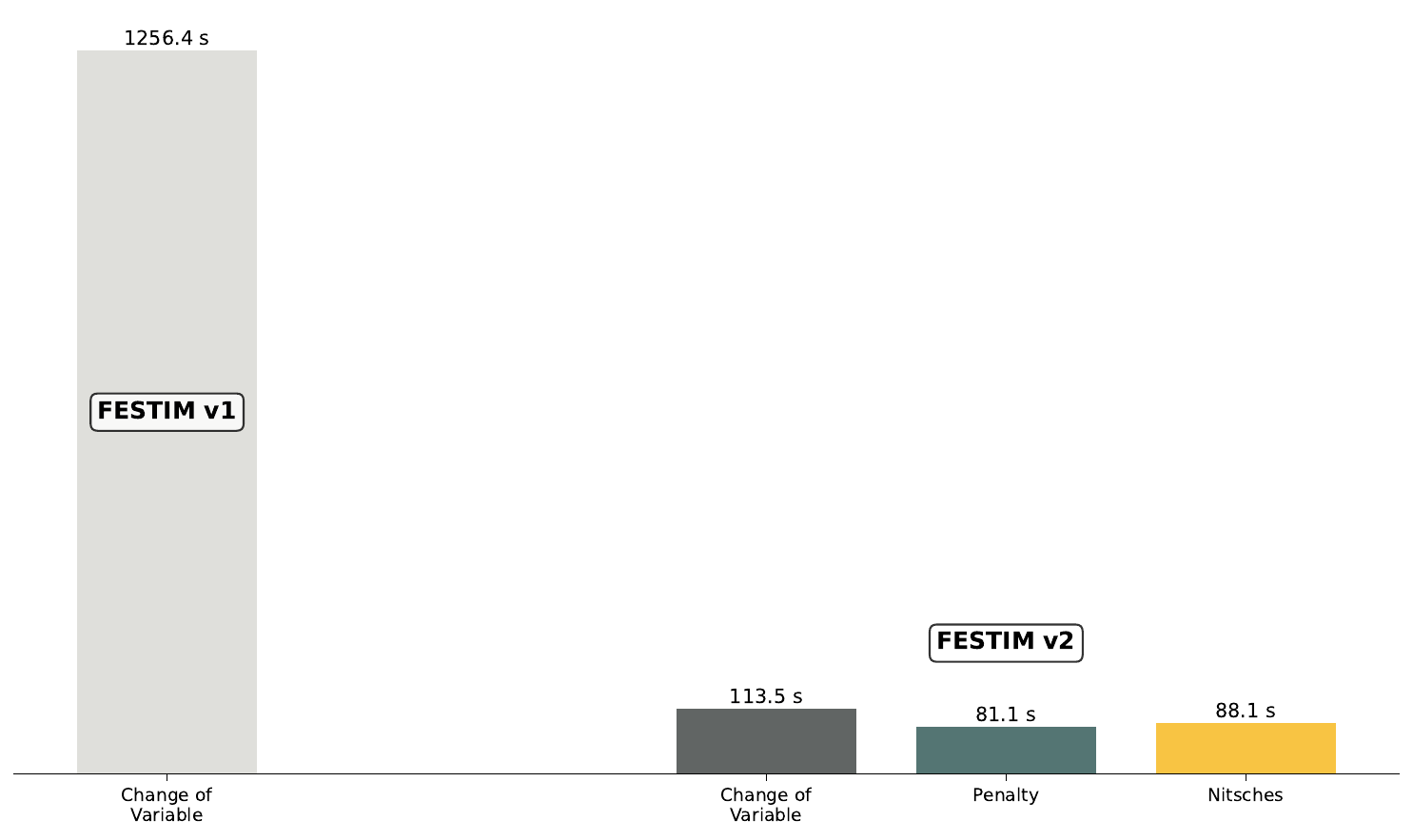}
    \caption{Comparison of average runtimes for the \textit{Diffusion: multi-material} benchmark case~\cite{delaporte-mathurin_festim_2024-1} using FESTIM v1 and v2.0 with different interface treatments. Results are averaged over ten runs.}
    \label{fig:performance_testing}
\end{figure}

FESTIM v1 required an average runtime of \SI{1256.4}{s} to complete the case (see \Cref{fig:performance_testing}). 
With the re-implemented change-of-variable method in FESTIM v2.0, the runtime was reduced to \SI{113.5}{s}, corresponding to a speed-up of about $11\times$. 
The discontinuous Galerkin approaches offered further improvements, achieving average runtimes of \SI{81.1}{s} with the penalty method and \SI{88.1}{s} with the Nitsche method, both around $15\times$ faster than FESTIM v1. 

These results demonstrate that FESTIM v2.0 delivers both greater flexibility and substantial improvements in computational performance. 
The change-of-variable method remains useful for continuity with legacy work but becomes less practical in large-scale, engineering-focused studies. 
By contrast, the discontinuous Galerkin formulations provide robust enforcement of interface conditions while reducing runtimes by almost an order of magnitude relative to FESTIM v1. 
As part of the ongoing development of FESTIM v2.0, a performance benchmark test suite will be established to systematically verify both numerical scaling and parallel scaling, in line with the design principles of DOLFINx~\cite{delaporte-mathurin_upgrading_2025}. 
For the present work, however, such parallel performance studies were outside the scope of this paper. 
Nevertheless, this step change in efficiency makes high-fidelity simulations of engineering-relevant components feasible, paving the way for multiphysics workflows, uncertainty quantification, and reactor-scale modelling with FESTIM v2.0.

\section{Multi-physics coupling}
\label{sec:multiphyics_coupling}
FESTIM's ethos in multi-physics modelling is to provide a robust and efficient hydrogen transport solver that integrates seamlessly with specialised external solvers, rather than reproducing their functionality. This approach prioritises interoperability, allowing FESTIM to be combined with established tools for heat transfer, fluid dynamics, and neutron transport.

To facilitate this interoperability, FESTIM v2.0 is accompanied by new companion packages developed within the FESTIM-dev ecosystem: \texttt{openmc2dolfinx} (\url{https://github.com/festim-dev/openmc2dolfinx}) and \texttt{foam2dolfinx} (\url{https://github.com/festim-dev/foam2dolfinx}). 
These libraries provide dedicated workflows for importing fields from OpenMC and OpenFOAM into the DOLFINx framework, enabling users to couple FESTIM directly with neutronics and CFD solvers.

\begin{figure}[h]
    \centering
    \includegraphics[width=0.75\linewidth, trim={6.75cm 5.25cm 6.75cm 4.25cm}, clip]{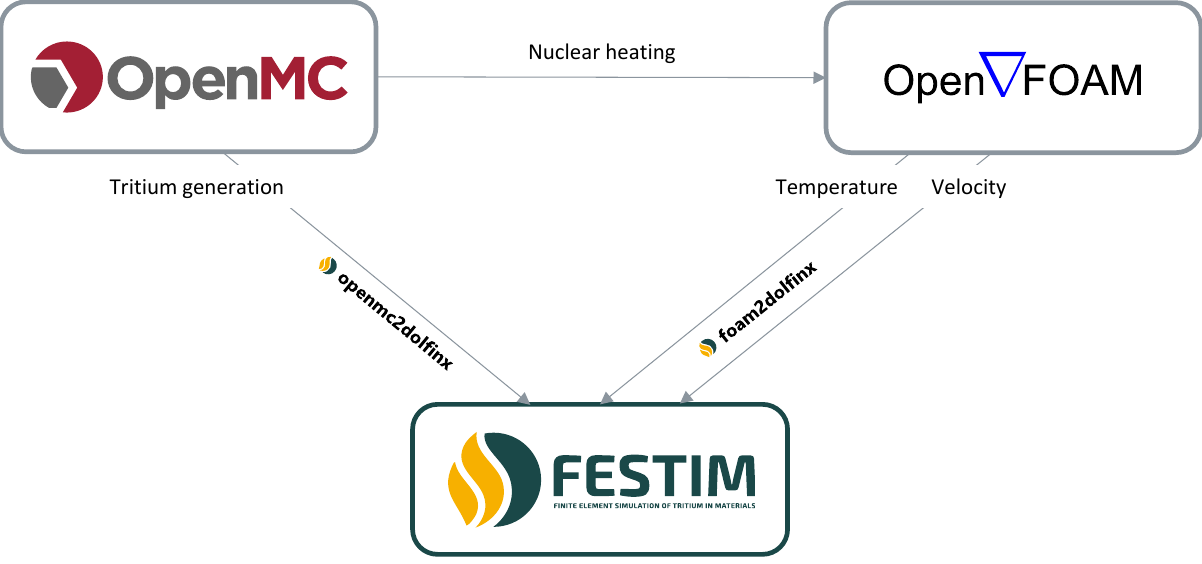}
    \caption{Fields passed between OpenMC, OpenFOAM, and FESTIM using the \texttt{openmc2dolfinx} and \texttt{foam2dolfinx} packages. This setup demonstrates FESTIM's extensibility and its ability to interoperate with specialised solvers in multi-physics workflows.}
    \label{fig:multiphsycs_coupling}
\end{figure}

\Cref{fig:multiphsycs_coupling} illustrates this philosophy: external codes generate fields such as temperature, velocity, or tritium generation, which are passed through the dedicated coupling tools into FESTIM. By focusing on extensibility and interoperability, FESTIM allows users to build multi-physics workflows tailored to their applications without compromising computational efficiency.

\subsection{Coupling CFD codes to FESTIM}
\label{subsec:cfd2festim}

FESTIM can incorporate velocity and temperature fields obtained from external CFD solvers to model advection-dominated hydrogen transport. 
However, CFD solvers such as OpenFOAM and finite element frameworks such as DOLFINx use different mesh and data formats, making direct interoperability non-trivial. 
To bridge this gap, the \texttt{foam2dolfinx} package was developed as part of the FESTIM-dev ecosystem. 
It enables FESTIM to use CFD fields directly by providing a lightweight interface between OpenFOAM and DOLFINx.

\subsubsection{foam2dolfinx}
The \texttt{foam2dolfinx} package streamlines FESTIM-OpenFOAM coupling by converting OpenFOAM fields into DOLFINx functions. 
It reads case files (\texttt{.foam}) and converts selected fields (e.g. velocity, temperature, turbulent viscosity) into \pyth{dolfinx.fem.Function} objects that can be used directly in FESTIM solvers. 

Internally, \texttt{foam2dolfinx} uses PyVista's \pyth{POpenFOAMReader}~\cite{sullivan_pyvista_2019} to parse OpenFOAM outputs and reconstruct an equivalent DOLFINx mesh by matching OpenFOAM cell connectivity to DOLFINx element topology. 
Supported cell types include tetrahedra (VTK type 10) and hexahedra (VTK type 12). 
This direct mapping of cell and vertex connectivity is significantly more performant than interpolation, while still preserving fidelity to the OpenFOAM solution. 
Fields can be taken at specified timesteps, and subdomains can also be targeted if required, allowing both steady-state and transient simulations to be handled naturally.

A minimal example is shown below:
\begin{python}
from foam2dolfinx import OpenFOAMReader
import festim as F

my_reader = OpenFOAMReader(filename="my_case.foam", cell_type=12)
velocity_field = my_reader.create_dolfinx_function(t=2.5, name="U")

...

F.AdvectionTerm(
    u=velocity_field
    ...
)
\end{python}

Here, the OpenFOAM reader loads the velocity field \texttt{U} at simulation time $t=\SI{2.5}{s}$ and converts it into a \pyth{dolfinx.fem.Function}.
This field is then passed directly into FESTIM's advection term, allowing seamless coupling between the two codes. 
In practice, this workflow is efficient and straightforward: OpenFOAM performs the fluid simulation, \texttt{foam2dolfinx} converts the relevant fields, and FESTIM natively uses them in hydrogen transport calculations.

\subsubsection{Lid-driven cavity example}
\begin{figure}[h!]
    \centering
    \begin{subfigure}{0.4\textwidth}
        \hspace*{-0.8cm}
        \includegraphics[width=\linewidth]{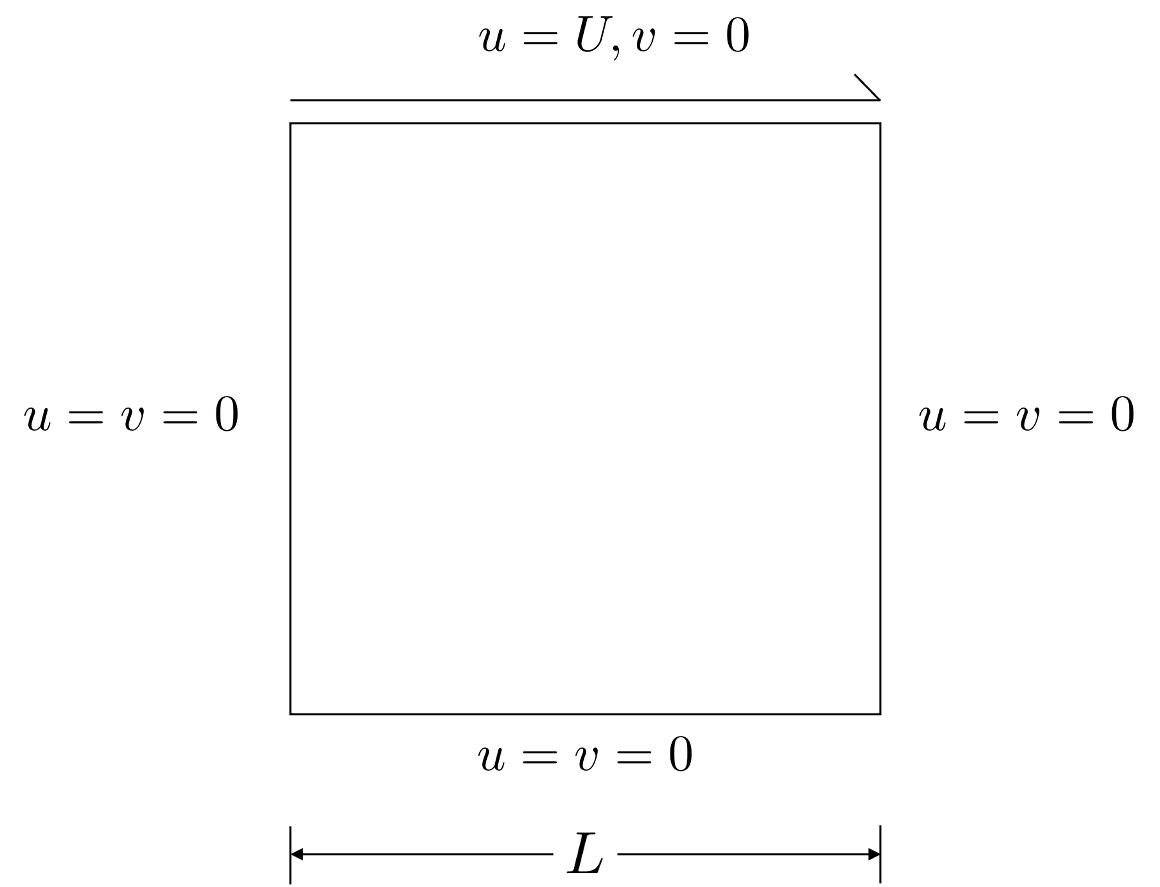}
        \caption{Lid-driven cavity problem setup}
        \label{fig:lid_driven_cavity}
    \end{subfigure}
    \quad \quad
    \begin{subfigure}{0.4\textwidth}
        \includegraphics[width=\linewidth]{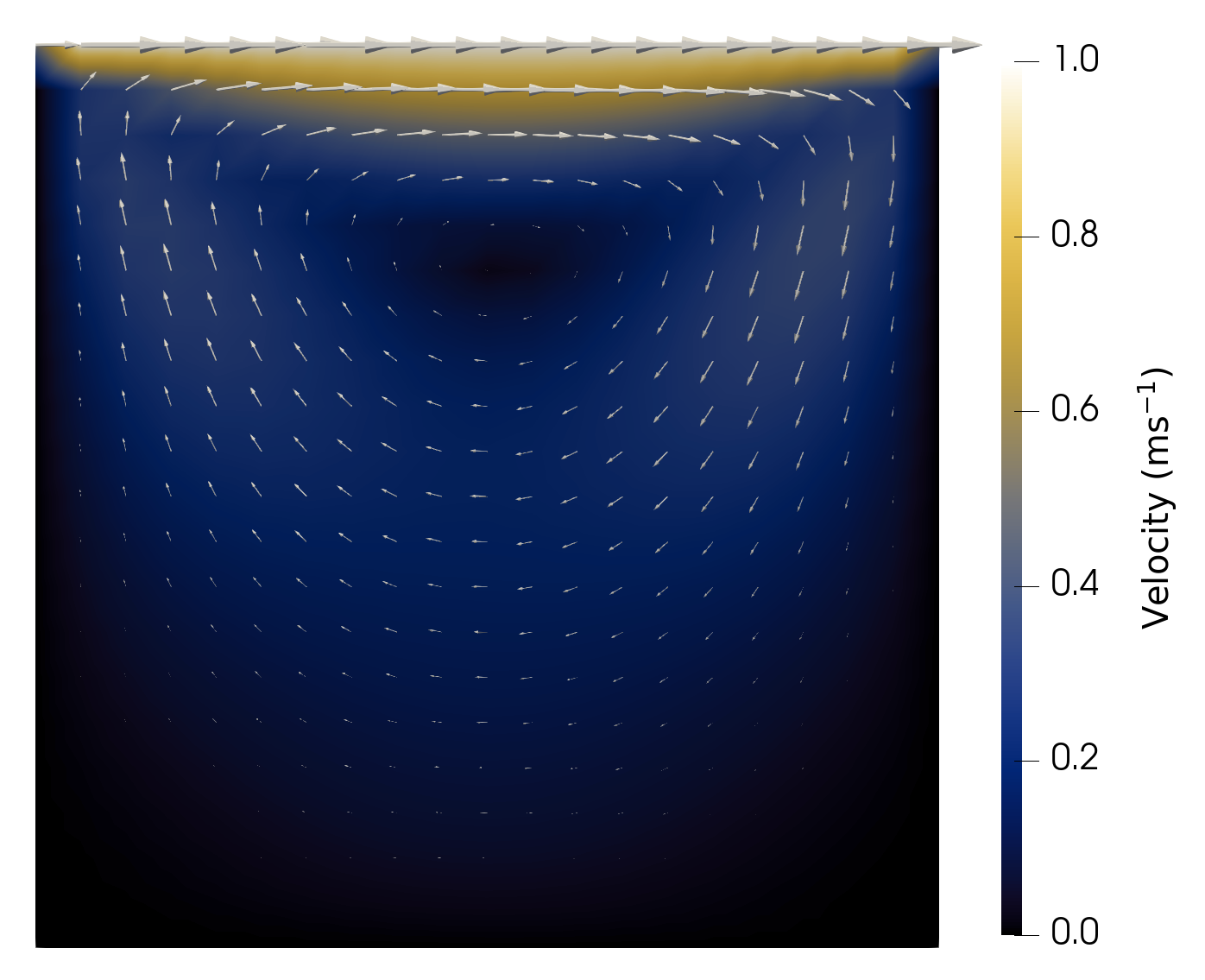}
        \caption{Velocity field from OpenFOAM}
        \label{fig:vel_field}
    \end{subfigure} \\[0.5cm]
    \begin{subfigure}{0.4\textwidth}
        \includegraphics[width=\linewidth]{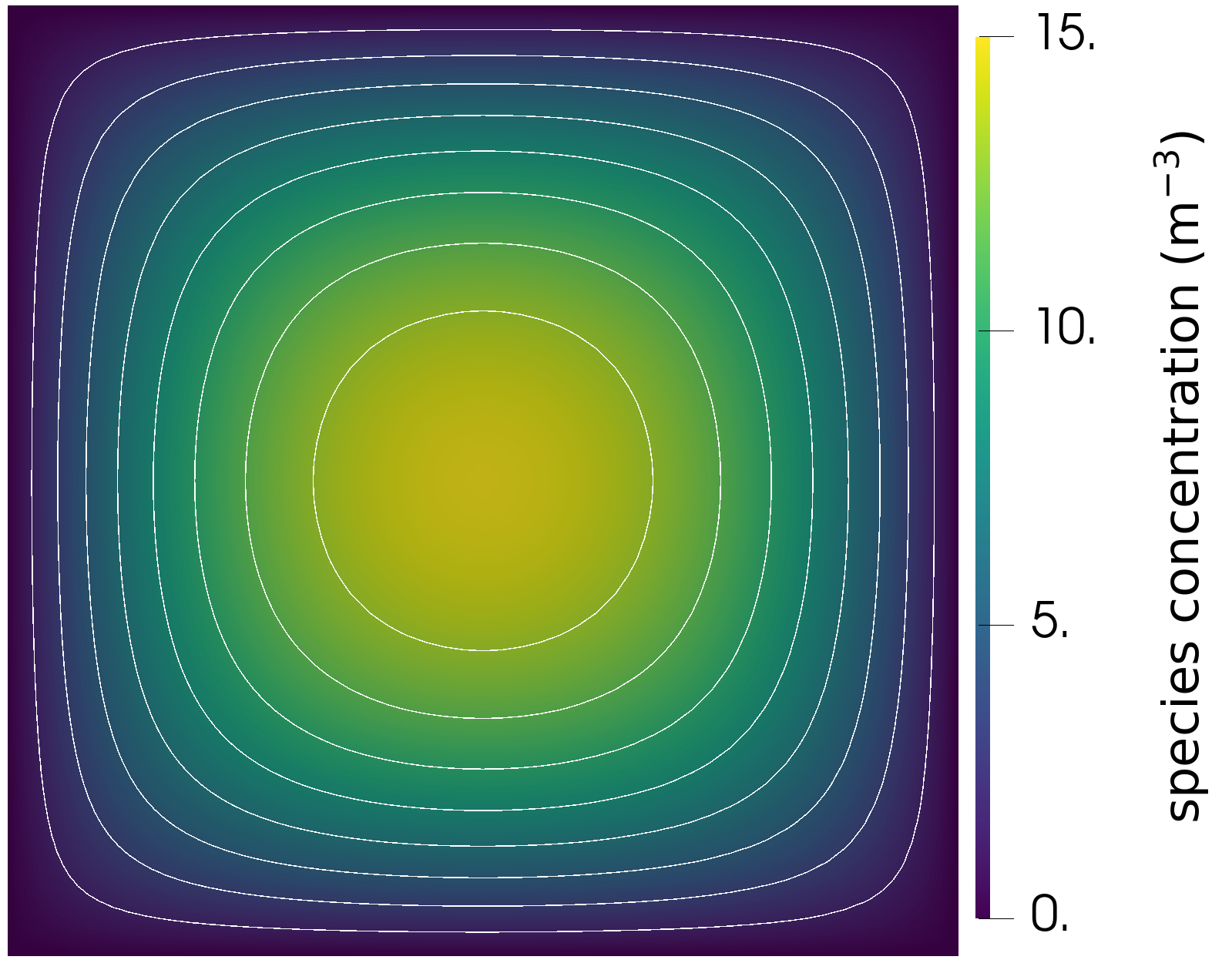}
        \caption{Species concentration without advection}
        \label{fig:ldc_no_advec}
    \end{subfigure}
    \quad \quad
    \begin{subfigure}{0.4\textwidth}
        \includegraphics[width=\linewidth]{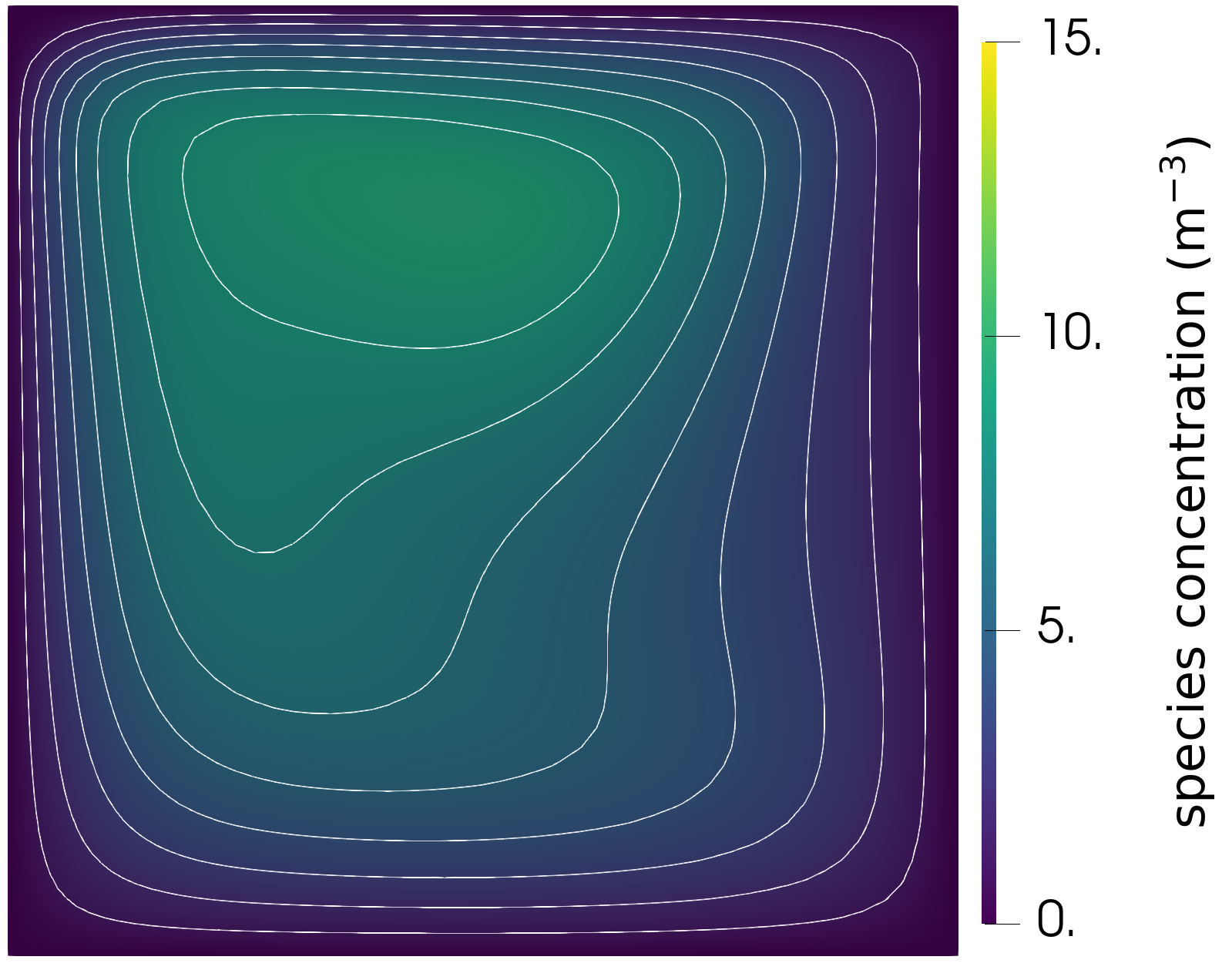}
        \caption{Species concentration with advection}
        \label{fig:ldc_advec}
    \end{subfigure}
        \caption{LDC benchmark: geometry, velocity field, and comparison of species concentration fields with and without advection.}
        \label{fig:ldc_results}
\end{figure}

The lid-driven cavity (LDC) setup consists of a square domain with a side length of $L = \SI{0.1}{m}$. 
No-slip boundary conditions are applied on the left, right, and bottom walls, while the top wall moves with a constant horizontal velocity $U = \SI{1}{m.s^{-1}}$. 
The problem is governed by the incompressible Navier-Stokes equations for a viscous fluid, solved in transient mode up to $t = \SI{2.5}{s}$. 
The velocity boundary conditions are:
\begin{subequations}
    \begin{align}
        \mathbf{u}(x=0, y) &= (0, 0), \quad
        \mathbf{u}(x=L, y) = (0, 0), \\
        \mathbf{u}(x, y=0) &= (0, 0), \quad
        \mathbf{u}(x, y=L) = (U, 0)
    \end{align}
\end{subequations}

where $\mathbf{u} = (u_x, u_y)$ is the velocity vector.
Once the velocity field is computed using OpenFOAM (see \cref{fig:vel_field}), it is exported and passed to FESTIM via \texttt{foam2dolfinx}.

For the FESTIM model, the domain is assumed to be filled with a fluid with a diffusion coefficient of $\SI{5e-4}{m^{2}.s^{-1}}$. 
A constant temperature of $\SI{500}{K}$ is imposed, together with a uniform volumetric source term of $\SI{10}{m^{-3}.s^{-1}}$.
The outer boundaries are set to a fixed concentration of zero, and the problem is solved in steady state. 
The mesh used for the hydrogen transport calculation is refined compared with that of the OpenFOAM simulation. 
Interpolation of the velocity field onto the FESTIM mesh is achieved using native functionality in DOLFINx, which efficiently supports non-matching meshes.

The concentration fields in \cref{fig:ldc_results} highlight the influence of advection on the transport process.  
Without advection, diffusion dominates, producing a symmetric profile with peak concentration at the centre and gradual decay to the zero-concentration boundaries (see \Cref{fig:ldc_no_advec}).  
With advection, species are transported more rapidly to the boundaries, and the concentration distribution follows the recirculating cavity flow (see \Cref{fig:ldc_advec}).  

This example demonstrates the FESTIM-OpenFOAM workflow. 
While the LDC problem is a simplified benchmark, it clearly illustrates how velocity fields can be incorporated into hydrogen transport calculations. With \texttt{foam2dolfinx}, the same approach extends naturally to more complex geometries and multi-physics scenarios, reinforcing FESTIM's extensibility and interoperability.

\subsection{Coupling neutronics codes to FESTIM}
\label{subsec:neutronics2festim}

In nuclear fusion systems, neutron and photon transport determine necessary source fields for hydrogen transport, most notably tritium generation and nuclear heating. 
Such fields are typically computed with Monte Carlo transport codes such as OpenMC~\cite{romano_openmc_2015}, which can tally spatial distributions of reaction rates and export the results in VTK format. 
The \texttt{openmc2dolfinx} package, developed within the FESTIM-dev ecosystem, provides a direct pathway for importing these tallies into FESTIM simulations. 
By converting OpenMC outputs into \pyth{dolfinx.fem.Function} objects enable nuclear source terms to be mapped seamlessly into hydrogen transport calculations without manual preprocessing.

\subsubsection{openmc2dolfinx}
The \texttt{openmc2dolfinx} package was developed to facilitate the transfer of OpenMC results into FESTIM. 
It reads \texttt{.vtk} files produced by OpenMC tallies and converts selected quantities (e.g.\ tritium generation rates, nuclear heating, or reaction scores) into \pyth{dolfinx.fem.Function} objects. 
These fields can then be used directly in FESTIM simulations.

Internally, the package relies on PyVista's \pyth{VTKDataSetReader} to parse OpenMC outputs. 
Because OpenMC tally values are constant across each mesh cell, the conversion process is more straightforward than for CFD fields: values are mapped cell by cell onto an equivalent DOLFINx mesh without requiring interpolation. 
Both structured (hexahedral) and unstructured (tetrahedral) tally meshes are supported via dedicated reader classes. 
This ensures that OpenMC results can be imported into FESTIM efficiently while preserving the fidelity of the transport solution.

A minimal example is shown below:
\begin{python}
from openmc2dolfinx import StructuredGridReader
import festim as F

reader = StructuredGridReader("tally_results.vtk")
t_gen_field = reader.create_dolfinx_function(data="(n,Xt)")

...

Tritium = F.Species("T")
my_source = F.Source(
    species=Tritium,
    value=t_gen_field
    ...
)
\end{python}

Here, the tritium generation tally \texttt{(n,Xt)} is read from an OpenMC \texttt{.vtk} file and converted into a \pyth{dolfinx.fem.Function}. 
This field is then passed directly to FESTIM as a source term for tritium transport.

\subsubsection{Lithium cube example}
To demonstrate the use of \texttt{openmc2dolfinx}, a benchmark was carried out on a simplified geometry: the irradiation of a lithium cube. 
This example highlights how neutron transport calculations from OpenMC can be imported into FESTIM and used for tritium transport.

\begin{figure}[h!]
    \centering
     \begin{subfigure}[b]{0.45\textwidth}
         \centering
         \includegraphics[width=\linewidth]{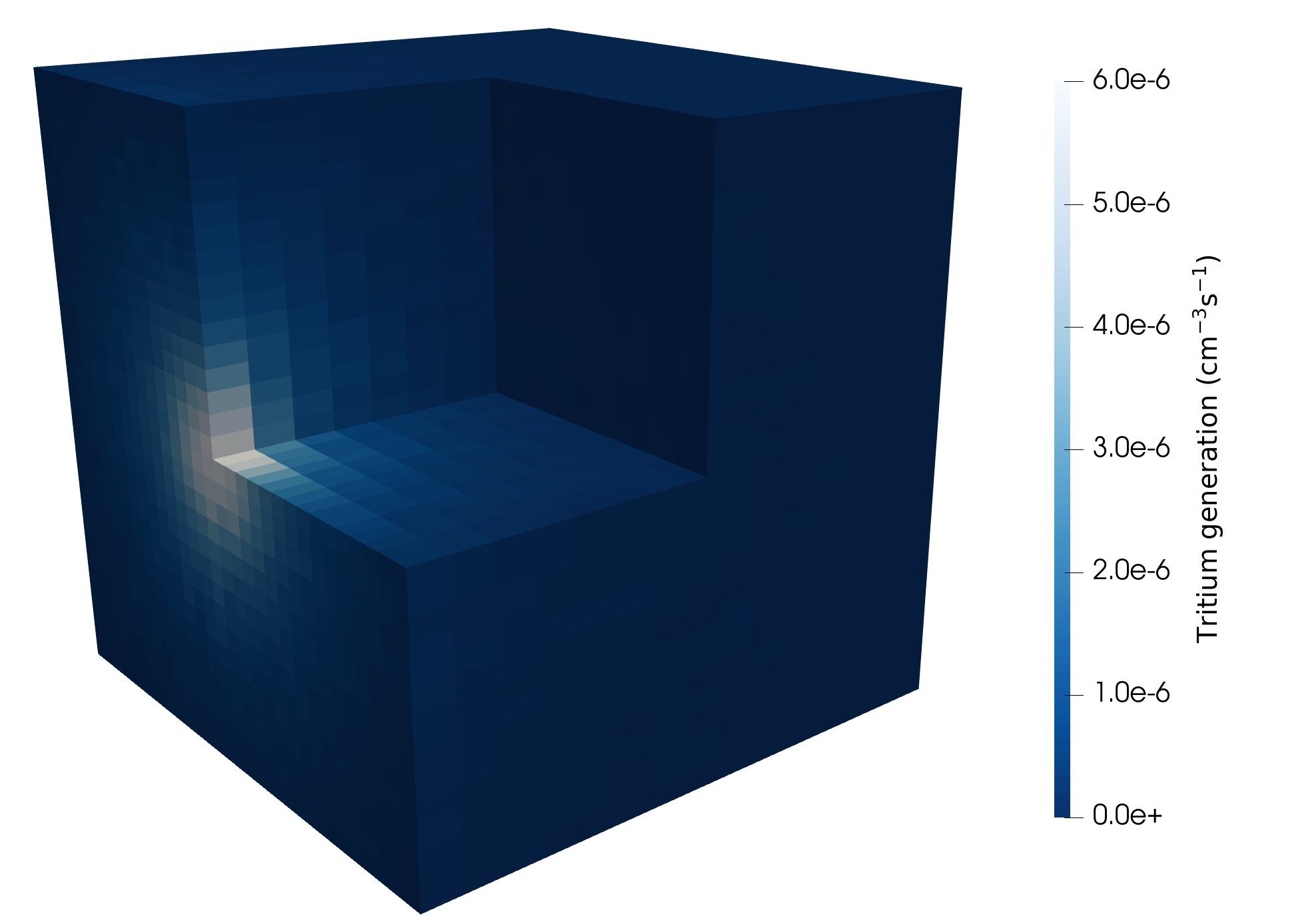}
         \caption{Tritium generation field from OpenMC}
     \end{subfigure}
     \quad \quad
     \begin{subfigure}[b]{0.45\textwidth}
         \centering
         \includegraphics[width=\linewidth]{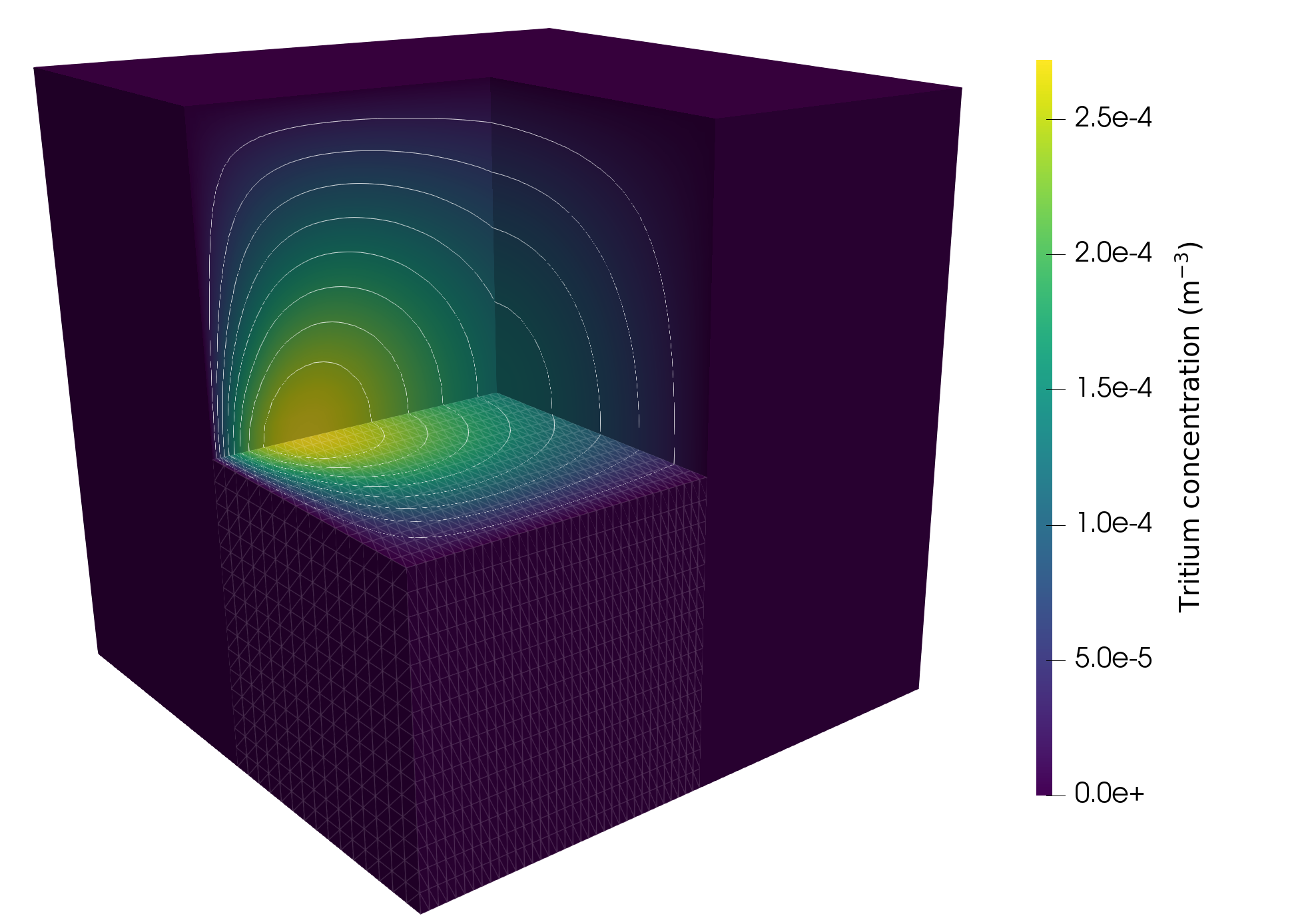}
         \caption{Steady-state tritium concentration from FESTIM}
     \end{subfigure}
        \caption{Coupling of OpenMC and FESTIM in the lithium cube example: tritium generation tallied in OpenMC (left) and resulting concentration field computed by FESTIM (right).}
        \label{fig:openmc_results}
\end{figure}

The OpenMC model consists of a lithium cube of side length \SI{120}{cm} surrounded by a vacuum boundary. 
A monoenergetic point source of \SI{14.1}{MeV} neutrons is placed \SI{10}{cm} outside the cube along the $z$ axis, emitting isotropically. 
The FENDL 3.1d nuclear data library is employed, and tritium generation is tallied using the \texttt{(n,Xt)} reaction score on a regular $30\times 30\times 15$ mesh. 
The resulting 3D tritium generation field is exported in VTK format.

In FESTIM, this field is imported using \texttt{openmc2dolfinx} and converted into a source term for tritium transport. 
The OpenMC tallies are produced on a structured hexahedral mesh. 
An equivalent hexahedral mesh is generated in \texttt{dolfinx}, which enables direct mapping of the tally values. 
In this example, the tritium transport is solved on a refined tetrahedral mesh that is denser than the original tally grid. 
\texttt{dolfinx} performs the interpolation between the hexahedral and tetrahedral discretisations, illustrating how fields can be transferred between meshes of different cell types (see \Cref{fig:openmc_results}). 
This interpolated source term is then used in the hydrogen transport simulation. 
The cube boundaries are set to zero concentration, the temperature is fixed at \SI{300}{K}, and the problem is solved in steady state.

This example illustrates the workflow for coupling OpenMC and FESTIM using \texttt{openmc2dolfinx}. 
While the lithium cube setup is simplified, it shows how nuclear source fields can be imported directly into FESTIM and used in hydrogen transport calculations. 
The same workflow extends to reactor-relevant cases, such as breeding blankets, where neutron fluxes govern tritium production and nuclear heating.

\section{Conclusions}
\label{sec:conclusions}
FESTIM v2.0, an open-source finite element framework for modelling hydrogen isotope transport in fusion-relevant materials, has been presented. 
The new release builds on earlier versions while introducing major extensions in scope and usability. 
The revised formulation adopts a modular structure, supporting multi-species transport and a wide range of physical processes, including multi-level trapping, isotope exchange, decay, and advection. 
Flexible interface and boundary condition treatments further extend the range of systems that can be represented within a single framework.

Beyond the physics, FESTIM v2.0 incorporates a refactored architecture designed for transparency, reproducibility, and ease of adoption. 
Improved documentation, version control, and continuous integration practices provide a sustainable foundation for community-driven development. 
Comprehensive export and post-processing options facilitate direct comparison with experiments and integration with multiphysics workflows.

Together, these advances establish FESTIM v2.0 as a versatile and sustainable platform for hydrogen transport studies in fusion materials. 
By lowering barriers to entry and providing a reliable computational backbone, the framework enables investigations ranging from fundamental science to engineering-scale tritium management. 
Future developments will focus on extending physical capabilities, broadening validation, and strengthening collaboration across the growing user community.

\section*{Acknowledgments}
The design of FESTIM v2.0 was guided by input from the user community.
A public discussion forum and monthly development meetings were established to facilitate collaboration, solicit feature requests, and support users. 
Community feedback highlighted the need for new capabilities, including multi-isotope transport, multi-level trapping, and streamlined coupling to external solvers. 
These requirements informed the development roadmap and design principles of the new version.

This material is based upon work supported by the National Science Foundation under Grant No. 2449339.

The authors would like to acknowledge funding from the Advanced Research Projects Agency-Energy (ARPA-E) of the U.S. Department of Energy. 
The views and opinions of authors expressed herein do not necessarily state or reflect those of the United States Government or any agency thereof.

\bibliographystyle{elsarticle-num}
\bibliography{references}

\end{document}